\documentclass[preprint2]{aastex6}
\usepackage{multirow}
\usepackage{amsmath}
\usepackage{stmaryrd}
\usepackage{graphicx}
\newcommand{\kevcm}{\text{\ keV\ cm$^2$}} 

\begin{document}

\title{pre-heating of the intergalactic medium by gravitational collapse and ultraviolet background}

\author{Weishan Zhu$^{1}$ and  Long-Long Feng$^{1,2}$}
\affil{$^{1}$Institute of Astronomy and Space Science, School of Physics and Astronomy, \\
Sun Yat-Sen University, Guangzhou 510275, China\\
$^{2}$Purple Mountain Observatory, CAS, Nanjing, 210008, China}

\begin{abstract}
The preheating of intergalactic medium by structure collapsing and ultraviolet background(UVB) are investigated in cosmological hydrodynamical simulations. When gravitational collapsing is the sole heating mechanism, we find that (1) $60\%, 45\%$ of the IGM are heated up to $S>8,  17$\kevcm \, respectively at $z=0$, but the fractions drop rapidly to a few percents at $z=2$; (2) the entropy of the circum-halo gas $S_{\rm{cir}}$ is higher than the virial entropy for more than $75 \%$ of the halos with masses $M<10^{11.5}$ $M_{\odot}$ since $z=2$, but the fraction higher than the entropy, $S_{\rm{pr}}$, required in preventive model of galaxies formation is only $15-20 \%$ for halos with $M<10^{10.5} M_{\odot}$ at $z=0$, and decreases as redshift increases; (3)assuming a metallicity of $Z \leq 0.03 Z_{\odot}$, the fraction of halos whose circum-halo gas having a cooling time longer than the Hubble time $t_{cool,cir}>t_{\rm{H}}$ is merely $5-10 \%$ at $z \lesssim 0.5$, and even less at $z \geq 1$ for halos with $M<10^{10.5} M_{\odot}$. (4) gas in the filaments undergoes the strongest preheating. Furthermore, we show that the UVB can not enhance the fraction of IGM with $S>17$\kevcm, but can increase the fraction of low mass halos($<10^{10.5} M_{\odot}$) that having $S_{\rm{cir}}>S_{\rm{pr}}$ to $\sim 70 \%$ at $z=0$, and that having $t_{\rm{cool, cir}}>t_{\rm{H}}$ to $15-30 \%$ at $z \lesssim 0.5$. Our results indicate that preheating due to gravitational collapsing and UVB are inadequate to fulfil the needs of preventative model, especially for halos with $10^{10.5}<M<10^{11.5} M_{\odot}$. Nevertheless, these two mechanisms might cause large scale galactic conformity.
\end{abstract}

\section{Introduction}
Modern galaxy formation and evolution models, with the aid of tools such as semi-analytical methods, cosmological N-body and hydrodynamical simulations, can reproduce the observations well regarding a wide range of galaxies properties(e.g., see recent review by Somerville \& Dave 2015 and reference therein). Despite the tremendous successes have been made, many details of important baryonic processes involved in galaxy evolution are not well resolved yet, such as the mechanisms that regulating the star formation activity(Naab \& Ostriker 2017). Currently, various internal feedback processes are implemented as the primary mechanism to tune the star formation, and hence fit the global cosmic star formation rate, as well as prevent the overcooling of gas in low-mass dark matter halos. The environment is also believed to have an important impact on the supply and cycling of gas in the formation and evolution of galaxies(e.g., Kauffmann et al. 2004; Baldry et al. 2006; Peng et al. 2010).

 Effects of large scale environment on the galaxies properties are probed in very recent observational studies(e.g., Alpaslan et al. 2016; Chen et al. 2017; Darvish et al. 2017; Kuutma, Tamm, \& Tempel 2017). The anisotropic gravitational collapse of structures in the universe was predicted to form web-like appearance of matter distribution on large scale, including voids, walls/sheets, filaments and clusters, currently referred to as the cosmic web(e.g., Zel'dovich et al. 1970; Icke et al. 1973; Bond et al. 1996). The observed galaxies distribution at low redshifts are basically consistent with this theoretical prediction(e.g., de Lapparent et al. 1986; Colless et al. 2003; Tegmark et al. 2004; Mehmet et al. 2014). Efforts have also been devoting to detect the intergalactic medium in the cosmic web, mostly the filaments, by different observational means(Cappetta et al. 2010; Chang et al. 2010; Cantalupo et al. 2014; Takeuchi et al. 2014). The heating and consequent cooling of gas during the collapse of structures will affect the accretion of gas in galaxies, and hence play an important role in their star formation history. The heating due to the virial shocks associated to dark matter halos have been intensively investigated and implemented in galaxy formation models(e.g., Binney 1977, White \& Rees 1978, White \& Frenk 1991). However, later studies suggested that filamentary cold streams could penetrate the shocks and feed galaxies in the dark matter halos that are less massive than a certain threshold(e.g., Dekel \& Birnboim 2006; Keres et al. 2009; ). Nevertheless, the more recent simulation work in Nelson et al(2013) showed that there were significant differences between different numerical solvers regarding the thermodynamic history of the accreted gas in halos more massive than $10^{10.5} M_{\odot}$. In addition to the virial shocks heating, multiply mechanisms could affect the thermal state of IGM and some would operate outside of the halos. 

The preheating of gas, i.e.,gas was heated up to certain temperature or entropy levels before collapse into dark matter halos, have been proposed to suppress the formation of low mass galaxies(Mo \& Mao 2002; Mo et al. 2005; Lu \& Mo 2007; Lu, Mo \& Wechsler 2015). Actually, UV photo heating has been shown to be able to prevent the collapse of baryons in halos up to $M=10^{10} h^{-1} M_{\odot}$ (e.g., Crain et al. 2007). Gravitational collapse of sheets and filaments can be additional potential mechanisms, as well as early star formation feedback, AGN, and perhaps dissipation of IGM turbulence(e.g., Mo \& Mao 2002; Lu \& Mo 2007; Zhu, Feng, \& Fang et al. 2011). Based on a semi-analytical model, Mo et al. (2005) demonstrated that the gravitational pre-heating due to the collapse of sheets at $z \lesssim 2.0$ can lift the entropy of circum-halo gas to $\sim 15$\kevcm. Using one-dimensional hydrodynamic simulation, Lu \& Mo(2007) further showed that preheated gas with entropy $S>8$\kevcm \, can strongly lessen the baryon accretion and gas cooling in haloes with masses $\mathbf{M}<10^{12} h^{-1} M_{\odot}$, if the mass accretion is smooth. Such level of preheating can lead to the mass of cooled gas scales with halo mass i.e., $M_{\rm{cool}} \propto M^2$, and hence match the observed atomic hydrogen and stellar mass function according to the model given in Mo et al.(2005). Recently, Lu et al.(2015) developed a new semi-analytic model of galaxy formation to probe the impact of preheating on disc galaxies, and showed that it can reproduce well a number of observational relations if the IGM was preheated up to a certain entropy level, i.e., $S\sim 17$\kevcm \, for halos with $M=10^{12} h^{-1} M_{\odot}$. 

However, the entropy state of intergalactic medium in filaments and other environments remains less tackled in three dimensional cosmological hydrodynamic simulations so far. The phase distribution of IGM in the temperature and density plane has been investigated thoroughly in simulations, focusing on the "missing" baryons at low redshifts(e.g., Fukugita, Hogan \& Peebles 1998; Fukugita 2004). The missing baryon is predicted to have temperature $10^5$ K$< T < 10^7$ K and moderate over-density in simulations,  commonly well known as ‘Warm Hot Intergalactic Medium(WHIM)’ (Cen \& Ostriker 1999; Dave et al. 2001; Cen \& Ostriker 2006), however,  that is still missing in the observations(e.g., Shull, Smith \& Danforth 2012). The link between gravitational heating of gas and star formation history has been examined only in a few works based on cosmological simulations. Cen(2011) argued that the gravitational heating of gas due to the growth of large scale structures and collapse of halos in over-dense regions on intermediate scales($\sim 1$Mpc) should be responsible for several global trends of galaxy evolution at low redshifts. The local mean entropy of gas at virial radius of halos in cluster region was found to be higher than those in void region in Cen(2011). Therefore, a continuous supply of cool gas to galaxies in over-dense regions was expected to end earlier than those in under-dense regions. However, the environments of filaments and sheets has not been fully probed. Based on N-body simulation, Liu \& Cen(2017) further proposed a model of gas accretion for galaxies, and suggested that gravitational shock heating of gas could be the primary cause that lead to the rapid drop of cosmic star formation rate at $z<2$. 

In the past decade, the evolution of cosmic web has been studied in detail by high resolution cosmological simulations(e.g.,Aragón-Calvo et al. 2007a; Hahn et al. 2007a; Aragon-Calvo et al. 2010 etc.), although most of these are pure N-body simulations. Attentions were paid to the alignment of filament with galaxies(e.g. Tempel et al. 2014;), the properties of the filaments, such as length, diameter and mass content(Cautun et al. 2014), and the impact of cosmic web on star formation(e.g., Snedden et al. 2016). It has been demonstrated that the dominant structures in term of mass fraction may transit from sheets to filaments at $z \sim 2-3$ for both baryonic and dark matter (Zhu \& Feng 2017). Meanwhile, the properties of cosmic shocks associated to structure formation have been also revealed by simulations, showing rapidly growth in strength and surface area since $z \sim 3$ (Ryu et al. 2003; Pfrommer et al. 2006; Kang et al. 2007; Skillman et al. 2008; Vazza et al. 2009, 2011; Zhu et al. 2013). Correspondingly, the entropy of intergalactic medium is expected to undergo a significant evolution. It would be interesting to make an investigation of the entropy level of gas heated by collapse of large scale structures and estimate their contributions to the entropy required in the preventative models. Moreover, the preheating of intergalactic medium might be the cause of galactic conformity extending out to up 4 Mpc in gas poor central galaxies with stellar mass $<10^{10} M_{\odot}$ (Kauffman et al. 2013,Kauffman 2015), although the significancy of the large scale conformity is under debate(e.g., Berti et al. 2017; Tinker et al. 2017). 

In this paper, we track the entropy of intergalactic medium along the evolution of cosmic web in three dimensional cosmological hydrodynamic simulations, and investigate corresponding impact of preheating. As a first step, the gravitational collapse and UV background are studied as the heating mechanisms in this work. This paper is organised as follows. We introduce the numerical methodology in Section 2. The entropy distribution of gas in different cosmic structures are investigated in Section 3. We than probe the preheating and cooling of gas within and surrounding dark matter halos in Section 4. In Section 5, we compare our results to works in the literature, estimate the power of gravitational collapse and UV background to fulfil the request by preventative models of galaxy formation, and discuss the numerical convergence, the impact of preheating on low mass galaxies and galactic conformity. Then we summarise our findings in Section 6. 

\section{Methodology}

\subsection{Simulations, and cosmic web classification}
To probe the effect of preheating by gravitational collapse, as well as UV background, we run two fixed grid cosmological hydrodynamic simulations with the Planck cosmology, i.e., $\Omega_{m}=0.317, \Omega_{\Lambda}=0.683,h=0.671,\sigma_{8}=0.834, \Omega_{b}=0.049$, and $n_{s}=0.962$(Planck Collaboration et al., 2014), in a periodic cubic box of side length $25 h^{-1}$ Mpc. One simulation is adiabatic, and the other one includes radiative cooling by assuming a pristine gas composition, i.e., $X=0.76, Y=0.24$. In the latter one, heating by a uniform UV background (Haardt \& Madau 2012, hereafter HM12) is switched on at $z=11.0$. The two simulations are referred to as 'L025-ada', and 'L025-uvc' respectively hereafter. The simulations are run by the hybrid N-body/hydrodynamic cosmological code WIGEON, in which the positivity-preserving Weighted Essentially Non-Oscillatory(WENO) scheme was implemented to solve the hydrodynamics(Feng et al. 2004; Zhu et al. 2013). The positivity-preserving WENO scheme can guarantee the positivity of density and temperature of gas, tackling the high mach problem without introducing any floor by hand(Zhang \& Shu 2012). All the simulations are evolved from redshift $z=99$ to $z=0$ in a $1024^3$ grid with equal number of dark matter particles. The simulations have space resolution $24.4 h^{-1}$kpc and mass resolution $1.30 \times 10^{6}M_{\odot}$. As a first step, we focus on the heating by gravitational collpase and UVB, processes such as the star formation, AGN and their feedback are not included. 

The classification of cosmic environments, namely, voids, sheets, filaments and knots is based on the tidal tensor of density field following Hanh et al.(2007b) and Forero-Romero et al.(2009). More specifically, the number of eigenvalues above a threshold value is used to tag the morphological type of environment that a cell belongs to. The threshold of eigenvalues is set to $0.8$. The same method and threshold were applied in Zhu \& Feng(2017), which showed that the dominant structures measured by mass fraction had transited from sheets to filaments at $z \sim 2-3$. 

\subsection{Gas entropy and dark matter halos}

The gas entropy is an excellent variable to estimate the gas cooling(Scannapieco \& Oh 2004). Following Mo et al. (2005), the specific entropy of gas is defined as, 
\begin{equation}
\begin{aligned}
S =\frac{T}{n_e^{2/3}} =17(\frac{\Omega_{b}h^2}{0.024})^{－2/3}(\frac{T}{10^{5.5}})\\
\times(\frac{1+\delta}{10})^{-2/3}(\frac{1+z}{3})^{-2} \kevcm ,
\end{aligned}
\end{equation}
where $\Omega_{b}$ is the cosmic density parameter at present, $\delta$ is the local overdensity, and the mean molecular weight is taken to $\mu=0.6$. 

The dark matter halos are identified using the FoF (Friend-of Friend) method, with a linking length parameter $0.2$. As the code is based on fixed grid, less massive halos are poorly resolved. We will investigate the preheating of gas for only those halos that consisting more than 250 dark matter particles in Section 4. The centre and radius of halos are defined by the sphere in which the mean density is 200 times of the critical density $\rho_{crit}(z)=3H^2(z)/8\pi G$ at $z$. The averaged thermal properties of gas within and surrounding the virial radius of halo, namely, halo gas and circum-halo gas,  are probed. We will carry out a brief convergence study on the resolution in Section 5. 

\section{Entropy of the IGM in various environments}

\begin{figure}[htbp]
\vspace{-1.5cm}
\hspace{-0.5cm}
\includegraphics[width=0.52\textwidth]{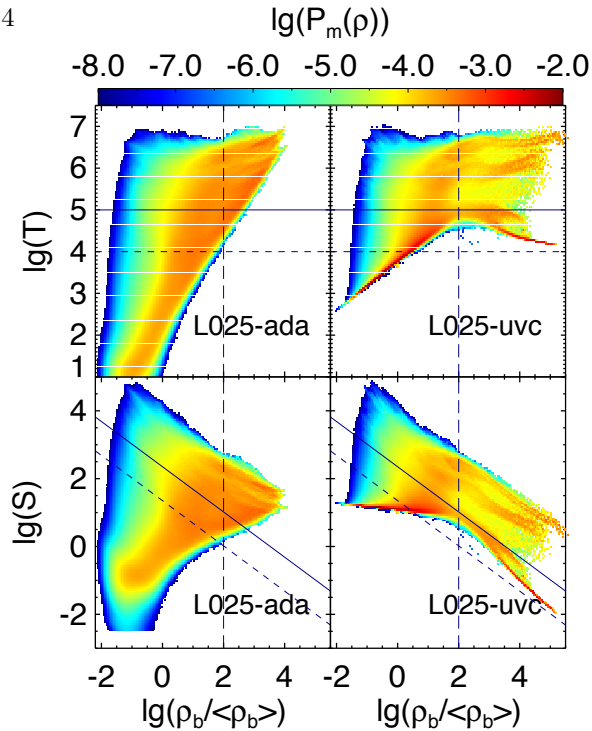}
\vspace{-1.5cm}
\caption{The mass weighted probability distribution of intergalactic medium in density-temperature(Top) and density-entropy(Bottom) plane at $z=0$. The left column shows the adiabatic run L025-ada and the right column shows the simulation with UV and radiative cooling (L025-uvc). Solid and short dashed lines indicate $T=10^5, 10^4$ K respectively. The long dashed line indicates $\rho_b/\bar{\rho}_b=100$}
\label{figure1}
\end{figure}

In this section, we investigate the entropy of the IGM in different cosmic structures, and their redshift evolution. 
Figure 1 compares the gas distribution in the density-temperature, and density-entropy space in our two simulations at $z=0$. 
The distributions in the density-temperature space are consistent with the current results in the literatures (e.g., Katz et al. 1996; Cen et al. 1999; Dave et al. 2001; Kang et al. 2007; Vazza et al. 2009). Adopting the definition of gas phases in Dave et al(2010) with the thresholds of $T_{th}=10^5$ K and $\delta_{th}=100$, most of the gas in the adiabatic simulation are in the phases of diffuse($T<T_{th}, \delta_b<\delta_{th}$), WHIM ($T>T_{th}, \delta_b<\delta_{th}$), and hot halo ($T>T_{th}, \delta_b>\delta_{th}$). The inclusion of UV heating and radiative cooling in the L025-uvc brings a slight change to the WHIM phase, but has a significant impact on the diffuse and hot halo gas. The UVB effectively heats up the diffuse IGM, and leads to a forbidden region in the temperature-density plane with a lower boundary of  $T(\rho_b) \propto T_0 (\rho_b/\bar{\rho}_b)^{\sim 0.6}$(e.g., Valageas et al. 2002; Vazza et al. 2009). The radiative cooling in highly over-dense regions promotes the collapse and results in a broader density range for the hot halo gas, and also helps to develop the cold condensed phase($T<T_{th}, \delta_b>\delta_{th}$).

Corresponding changes in the density-entropy phase plane can be observed in the two bottom panels of Figure 1, reflecting the transformation according to eqn(1). For over dense regions of gas with $\delta_b>100$, the entropy of gas in L025-uvc shows an evident decrease in comparison to L025-ada due to cooling. Gas with $\delta_b>100$ would had been heated up to $T> 10^{4.5}$ K by gravitational collapse and the UVB, at which the cooling will be effective and immediately lead to the decrease of entropy in L025-uvc. With the lower boundary in the temperature-density plane, i.e., $T(\rho_b) \sim 10^{3.5} (\rho_b/\bar{\rho}_b)^{0.6}$, the corresponding entropy floor reads $S \sim 13(\rho_b/\bar{\rho}_b)^{-0.07}$\kevcm \ at $z=0$, as shown by the bottom right panel of Figure 1. On the other hand, the WHIM gas would undergo mild changes in response to the UV ionizing background. The thermal evolution of the WHIM gas should be governed by gravitational heating. 
\begin{figure*}[htbp]
\vspace{-2.8cm}
\gridline{
\hspace{-2.5cm}
\includegraphics[width=0.40\textwidth]{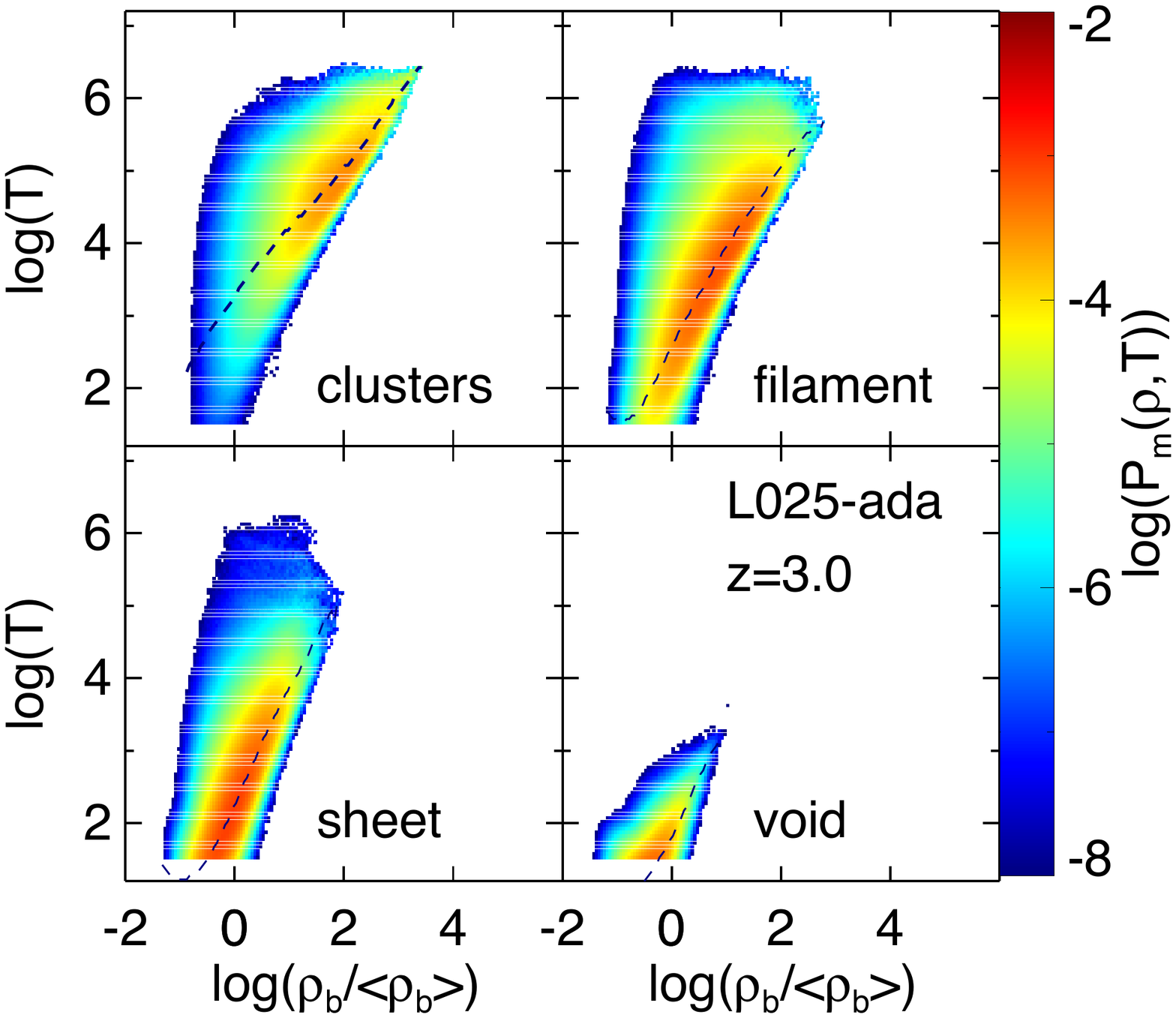}
\hspace{-3.0cm}
\includegraphics[width=0.40\textwidth]{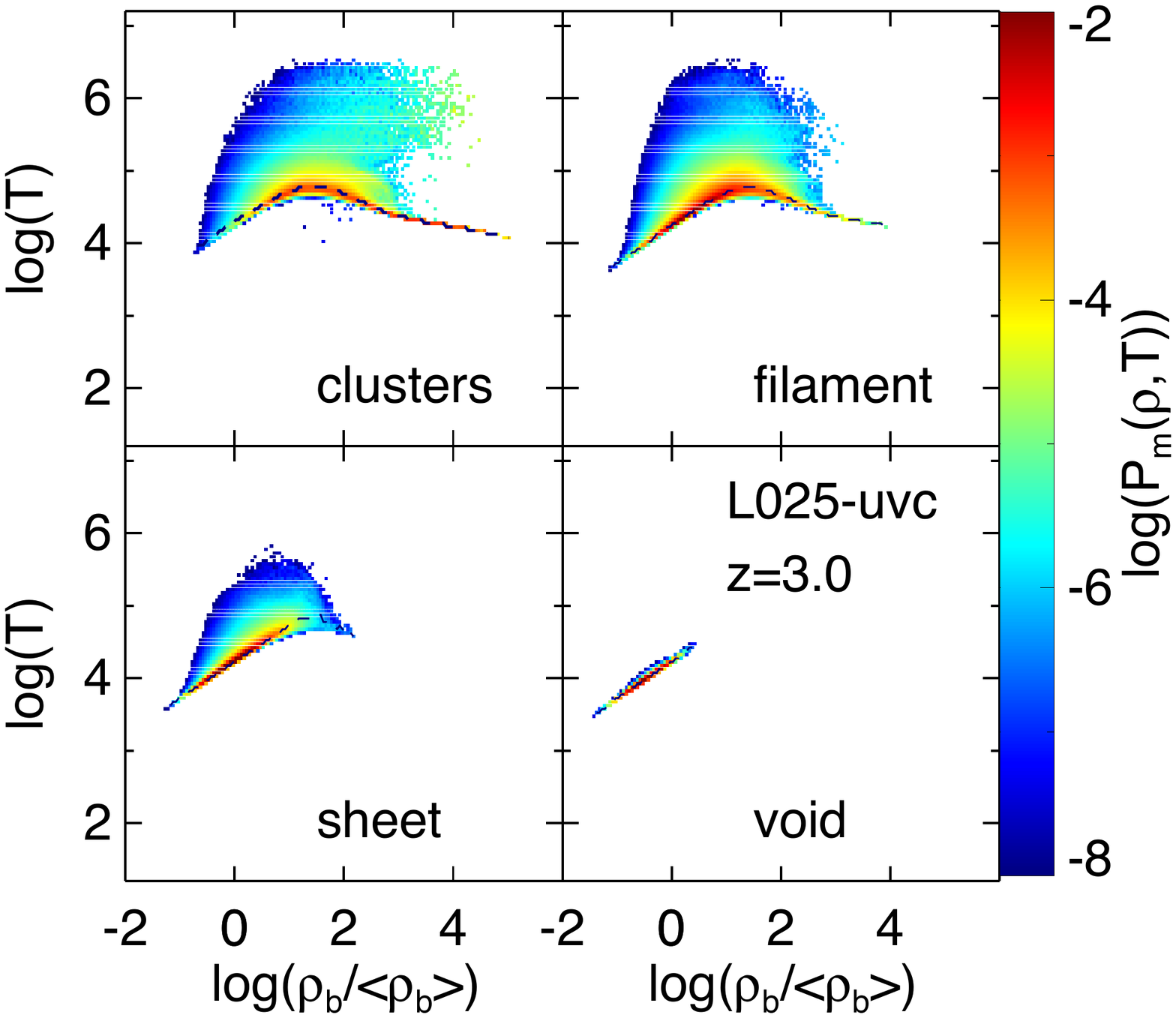}
}
\vspace{-4.0cm}
\gridline{
\hspace{-2.5cm}
\includegraphics[width=0.40\textwidth]{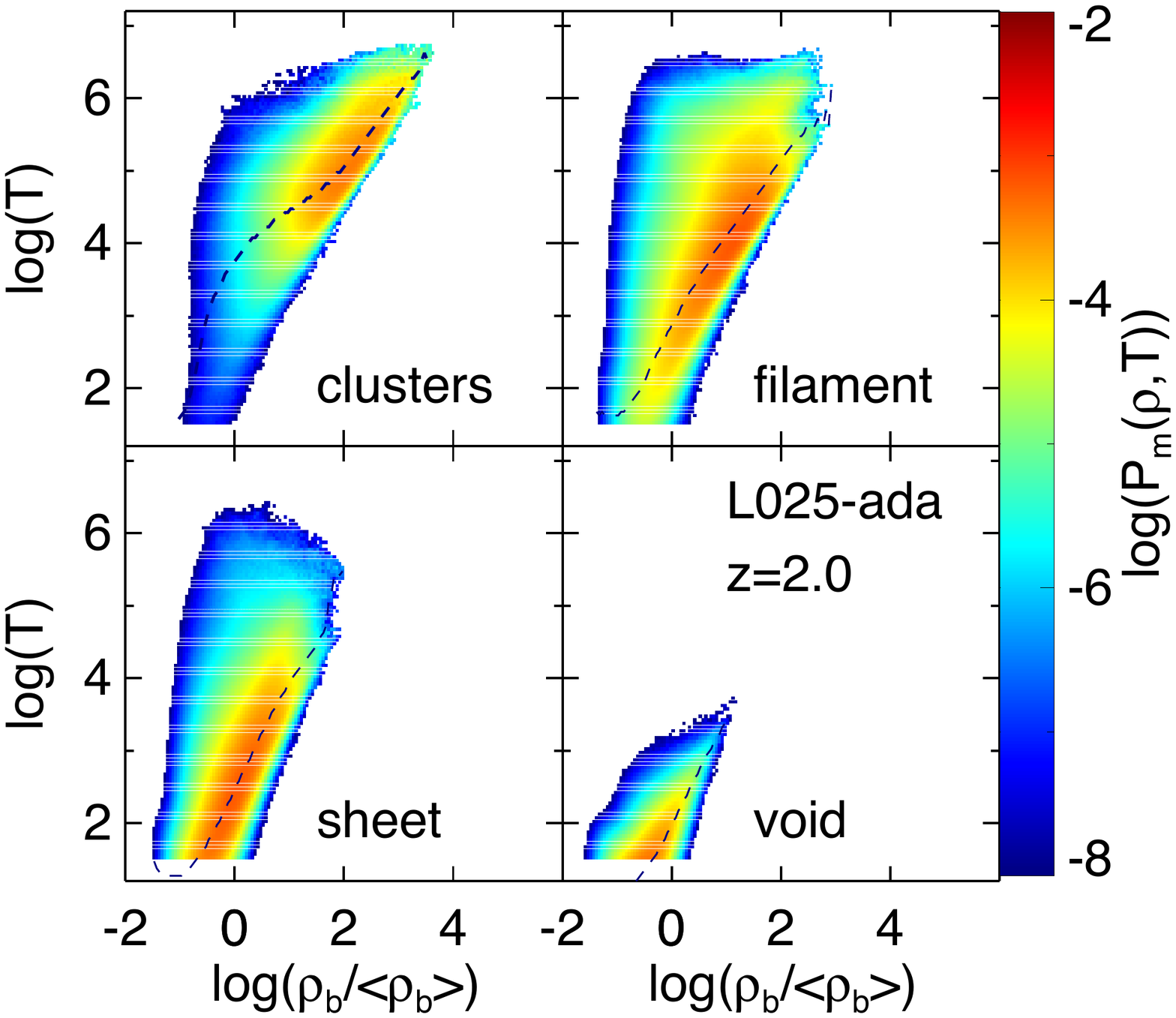}
\hspace{-3.0cm}
\includegraphics[width=0.40\textwidth]{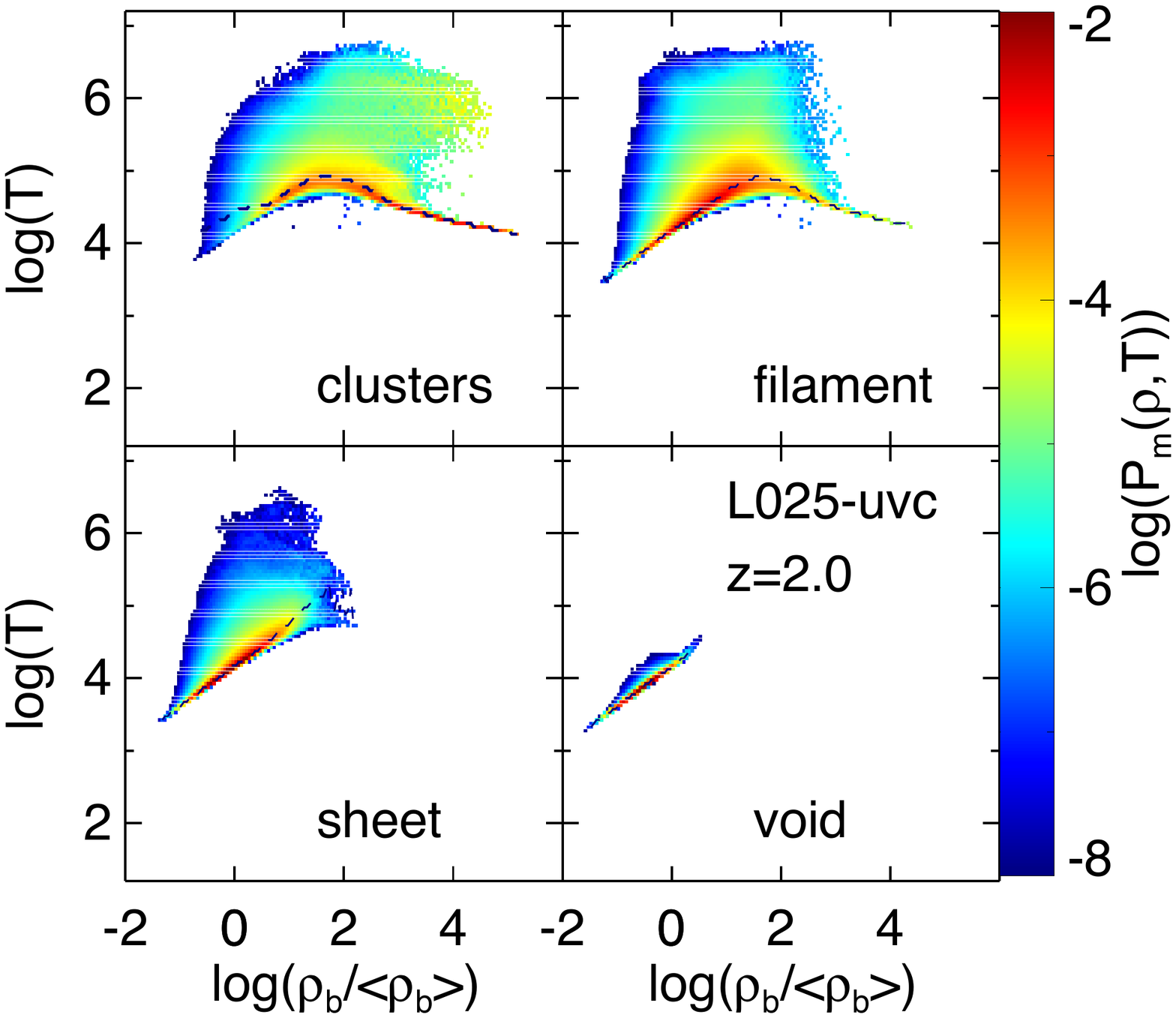}
}
\vspace{-4.0cm}
\gridline{
\hspace{-2.5cm}
\includegraphics[width=0.40\textwidth]{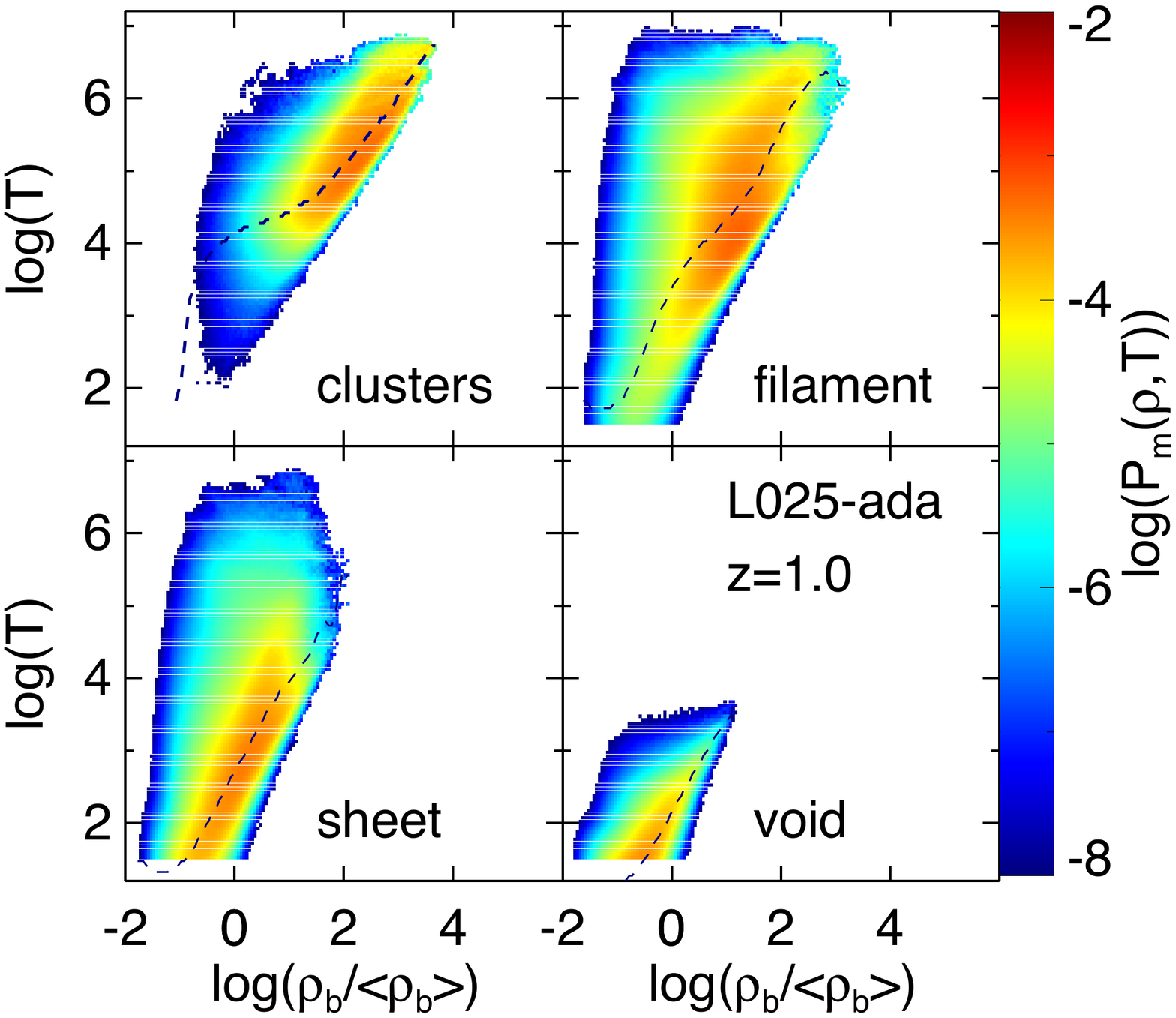}
\hspace{-3.0cm}
\includegraphics[width=0.40\textwidth]{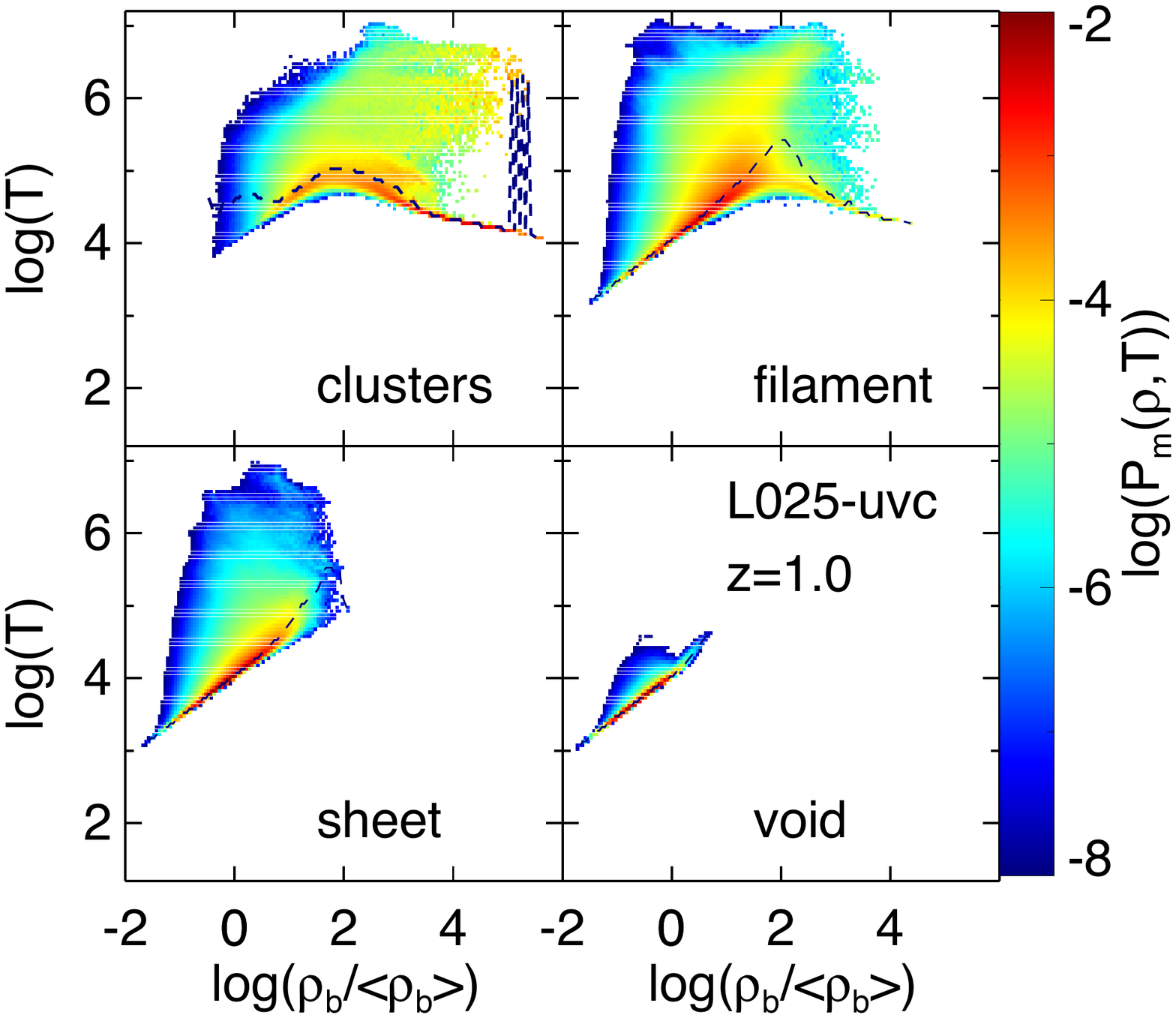}
}
\vspace{-4.0cm}
\gridline{
\hspace{-2.5cm}
\includegraphics[width=0.40\textwidth]{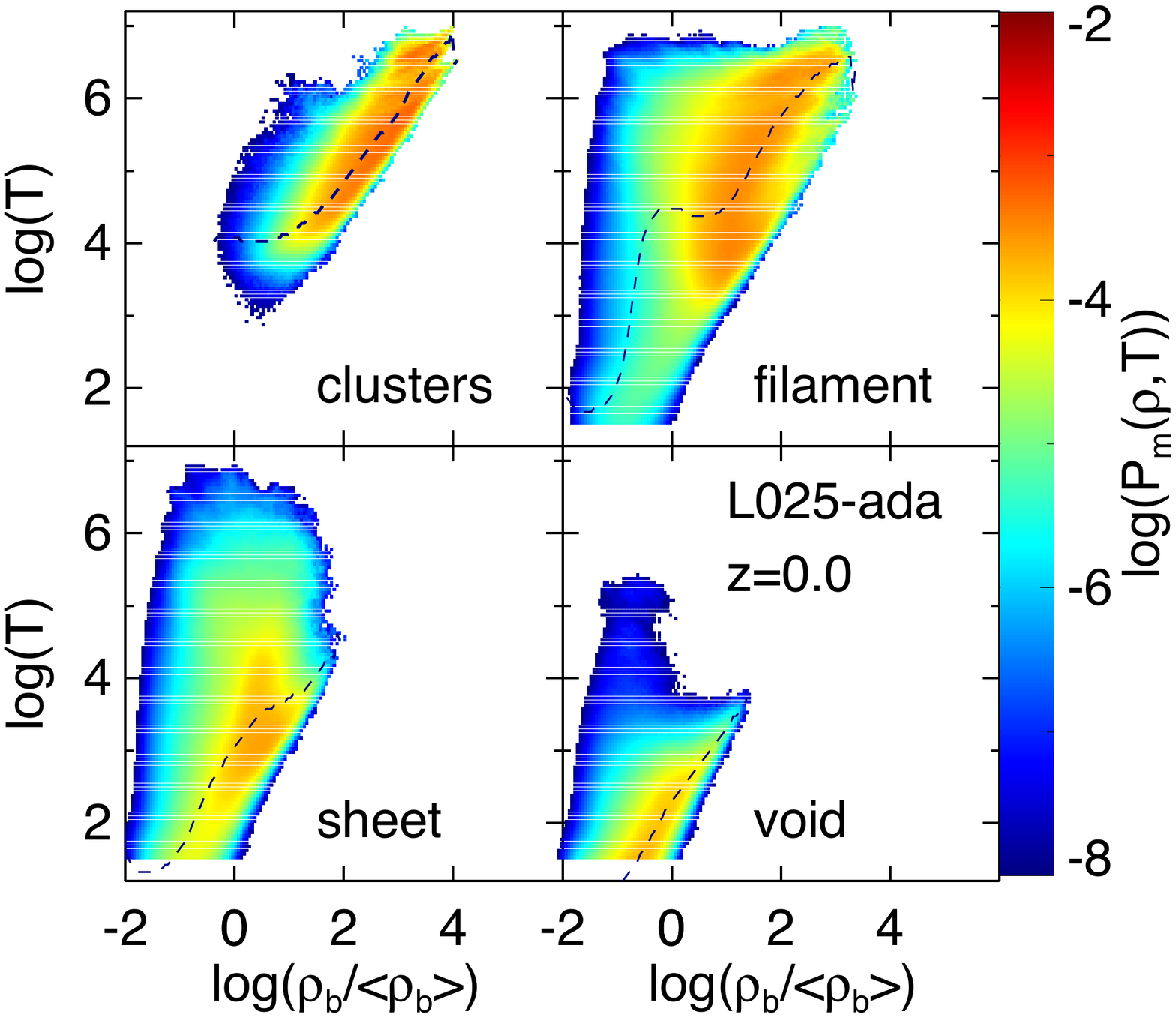}
\hspace{-3.0cm}
\includegraphics[width=0.40\textwidth]{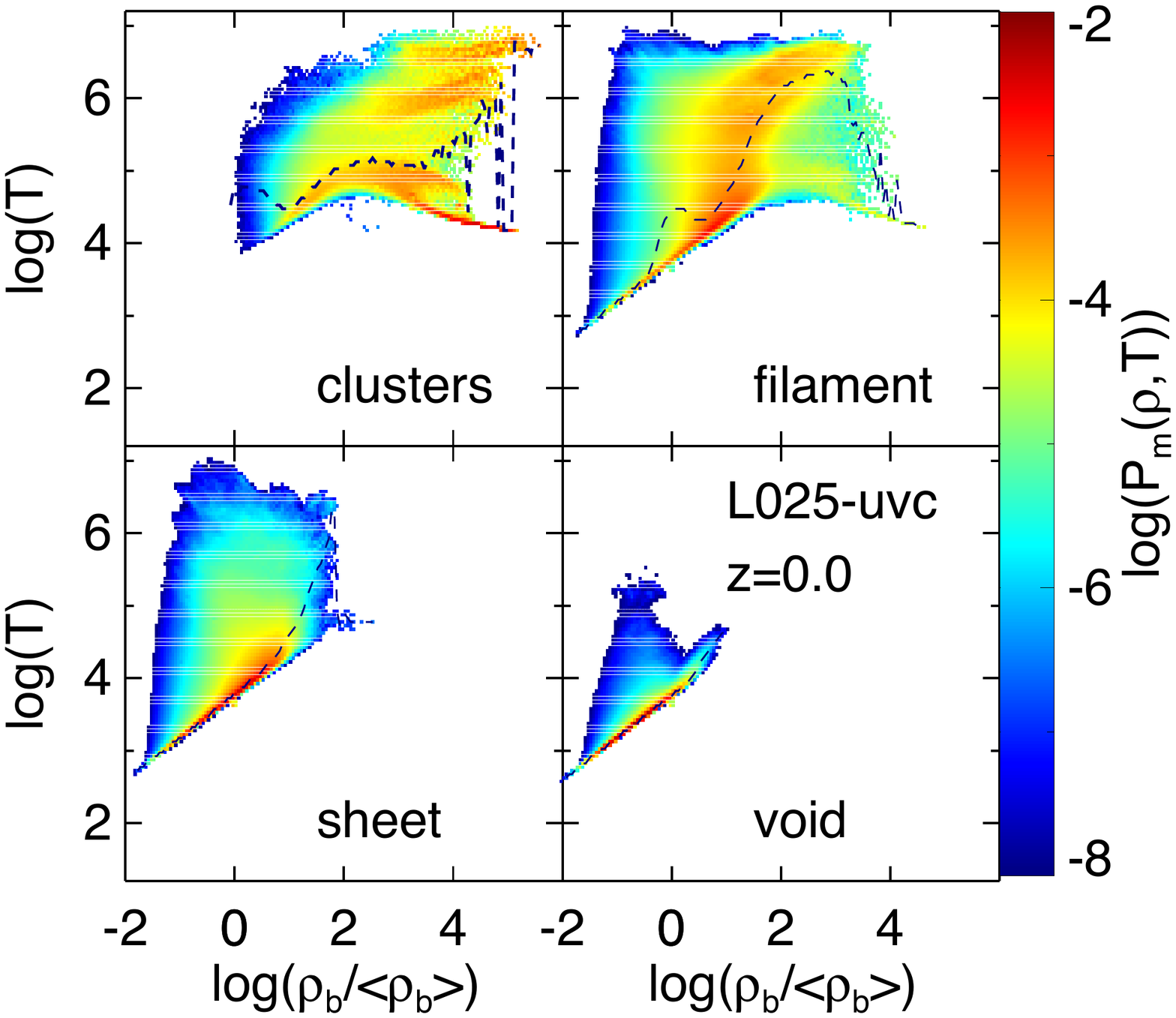}
}
\vspace{-2.0cm}
\caption{The mass weighted distribution in the density-temperature phase diagram of intergalactic medium residing in different environments in simulation L025-ada(Left) and L025-uvc(Right) at $z= 3.0, 2.0, 1.0, 0.0$(from top to bottom). Dashed curves indicate the median values of gas temperature in the corresponding density bin.}
\label{figure2}
\end{figure*}

\begin{figure*}[htbp]
\vspace{-2.8cm}
\gridline{
\hspace{-2.5cm}
\includegraphics[width=0.40\textwidth]{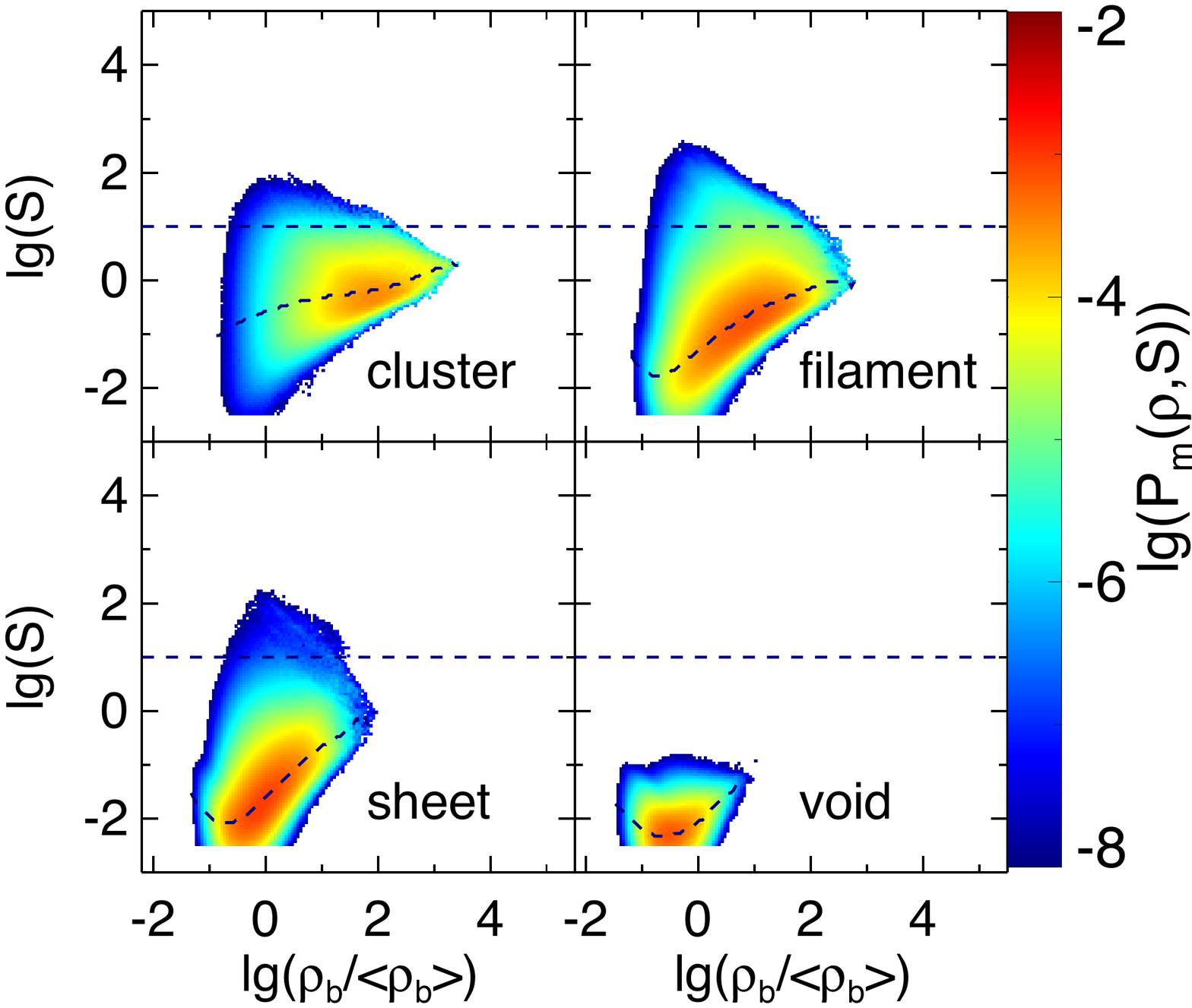}
\hspace{-3.0cm}
\includegraphics[width=0.40\textwidth]{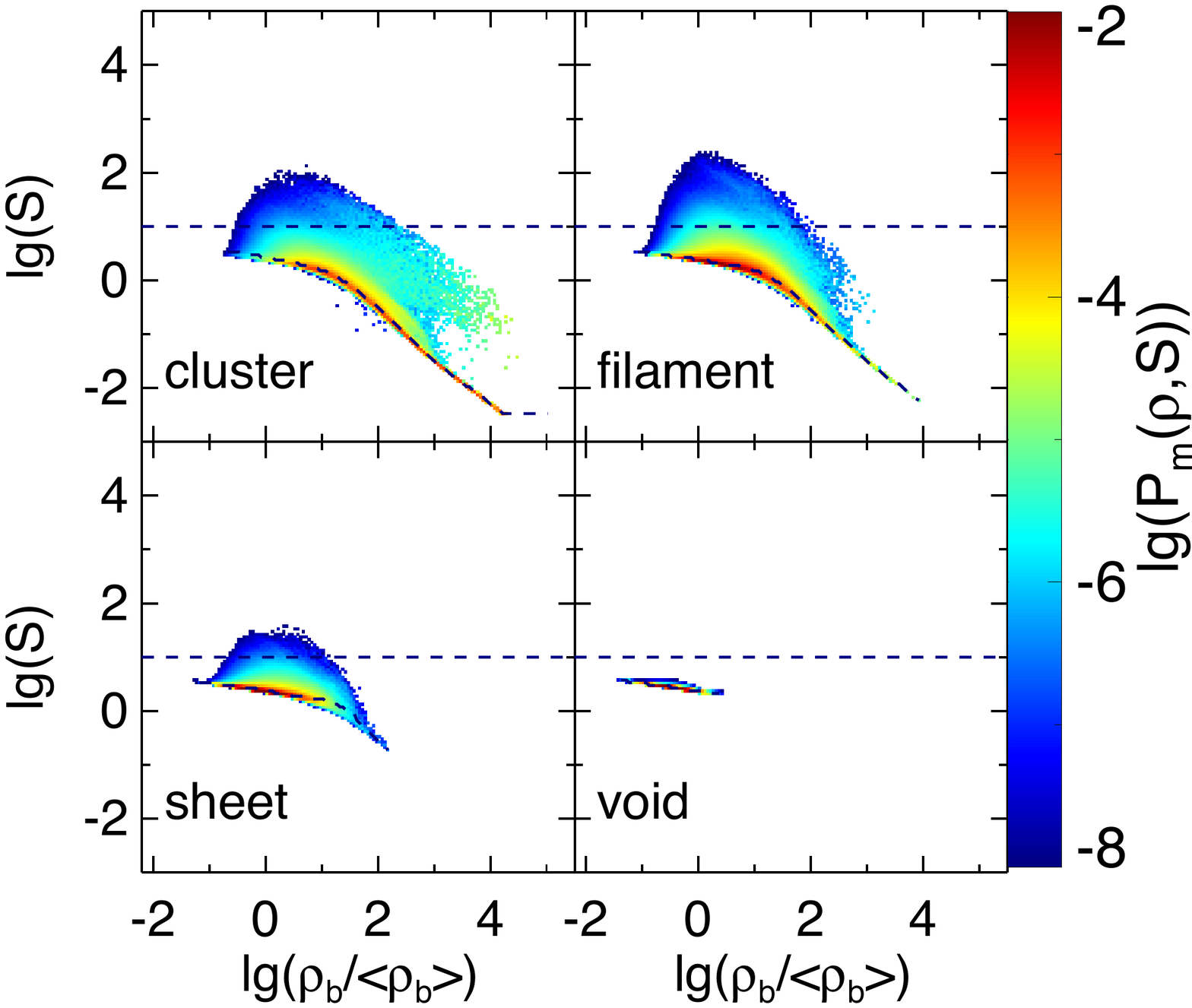}
}
\vspace{-4.0cm}
\gridline{
\hspace{-2.5cm}
\includegraphics[width=0.40\textwidth]{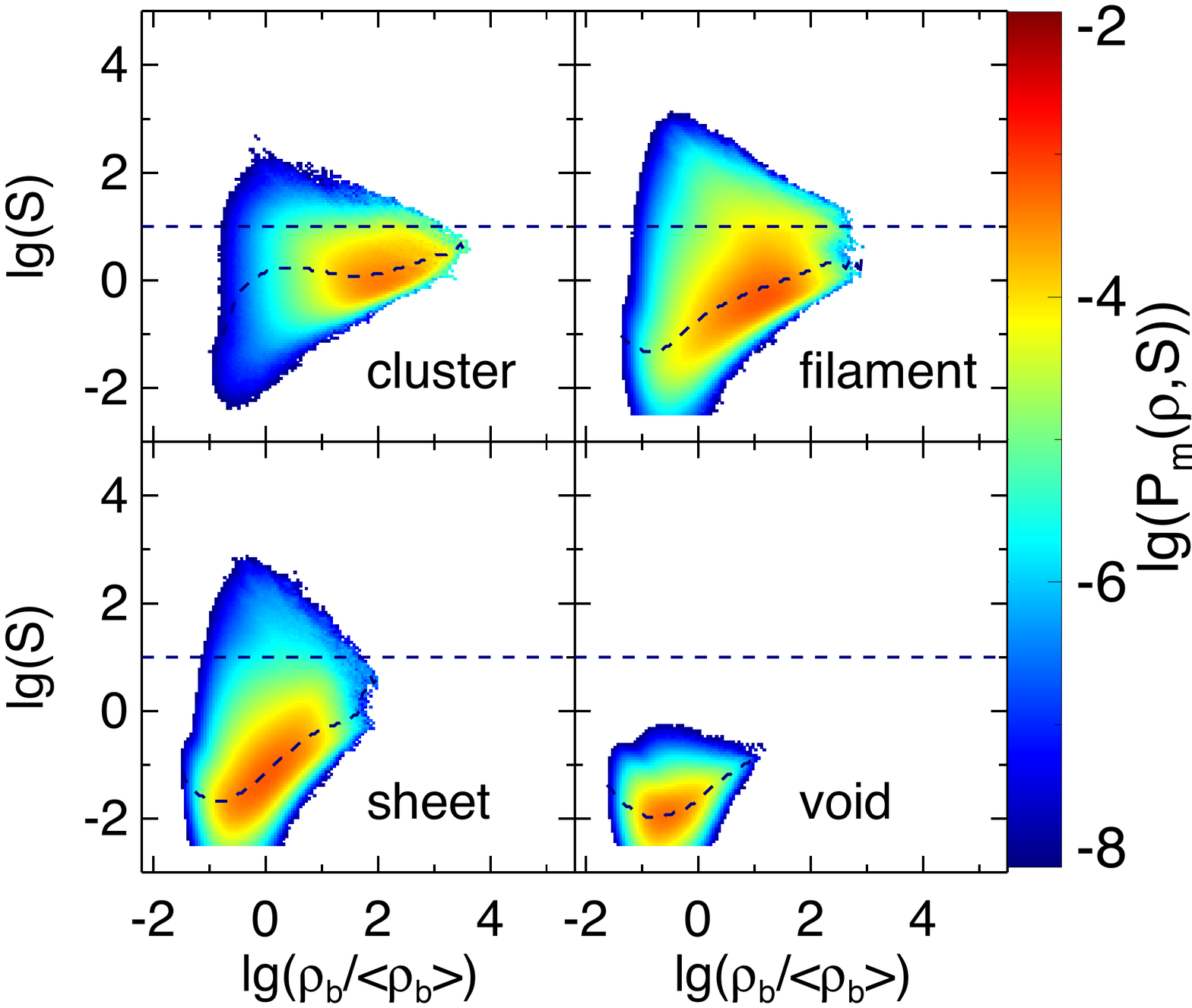}
\hspace{-3.0cm}
\includegraphics[width=0.40\textwidth]{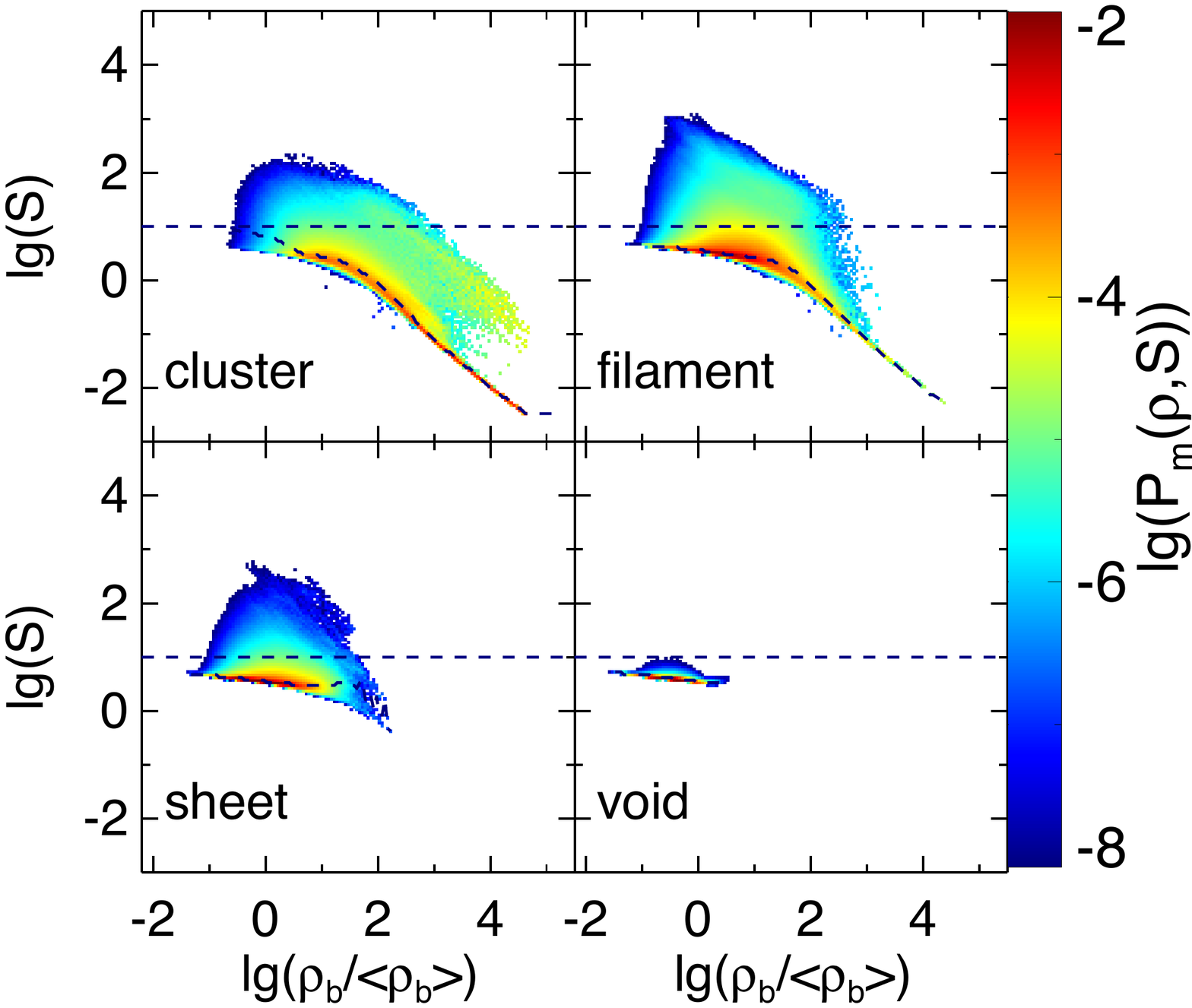}
}
\vspace{-4.0cm}
\gridline{
\hspace{-2.5cm}
\includegraphics[width=0.40\textwidth]{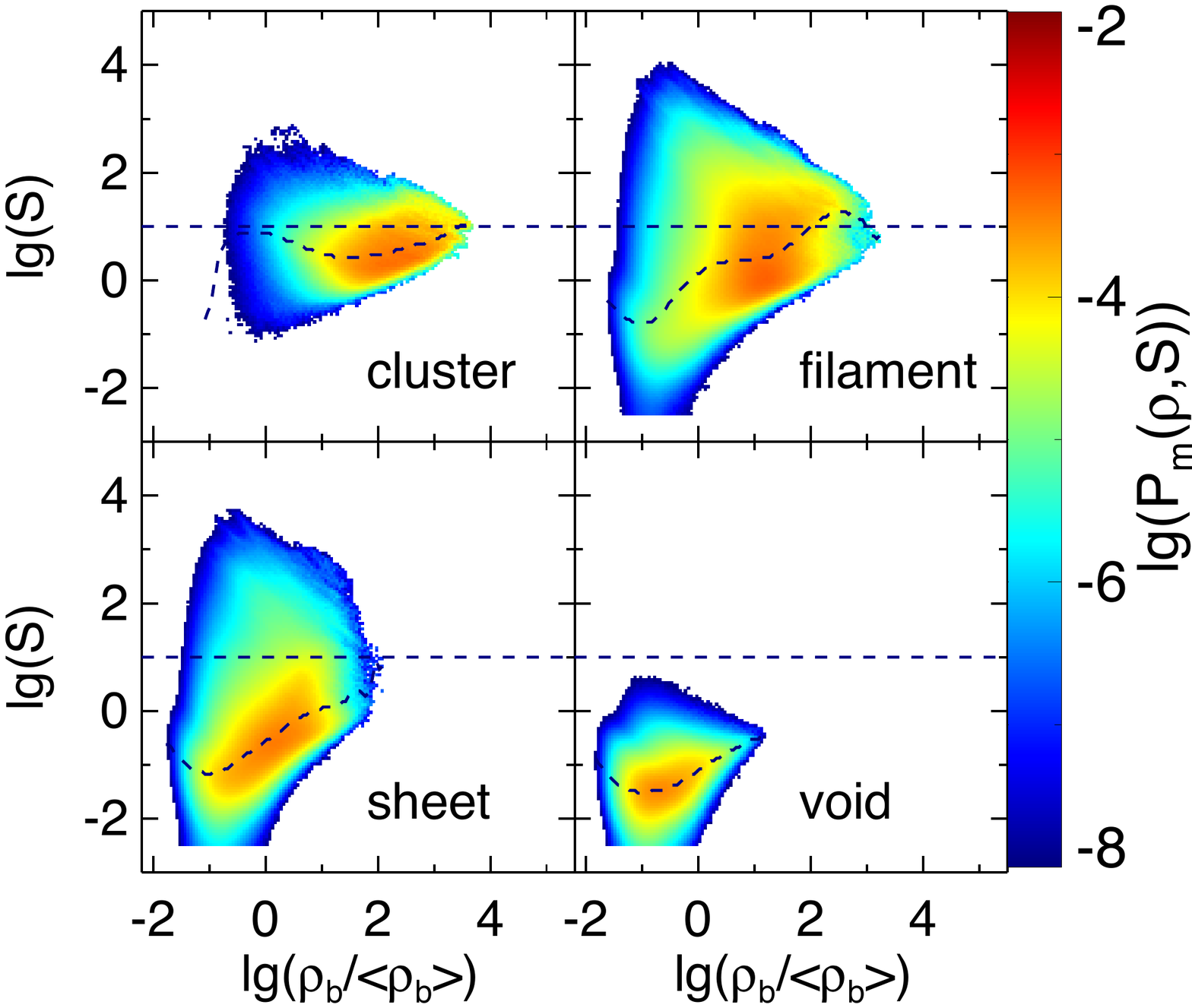}
\hspace{-3.0cm}
\includegraphics[width=0.40\textwidth]{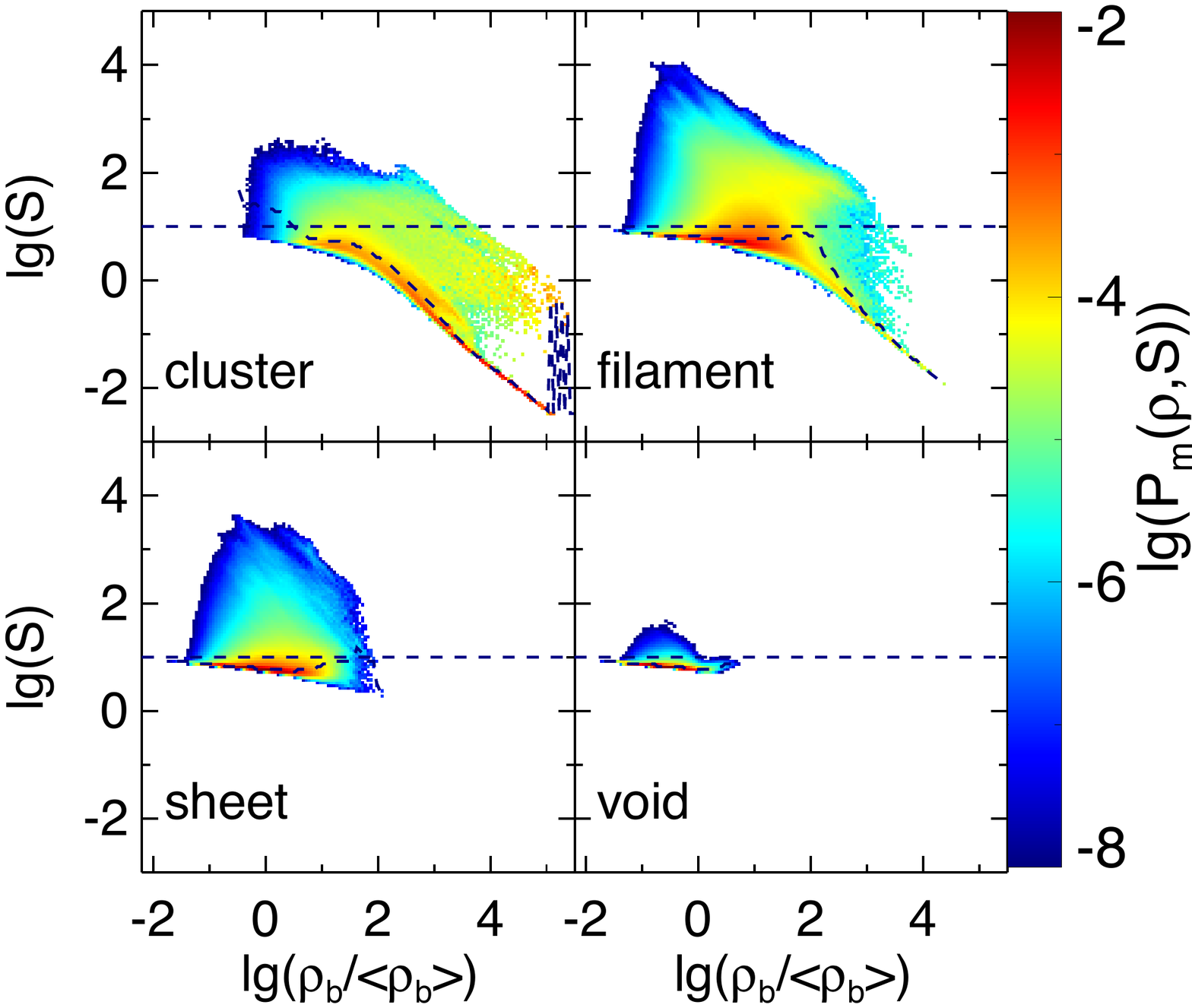}
}
\vspace{-4.0cm}
\gridline{
\hspace{-2.5cm}
\includegraphics[width=0.40\textwidth]{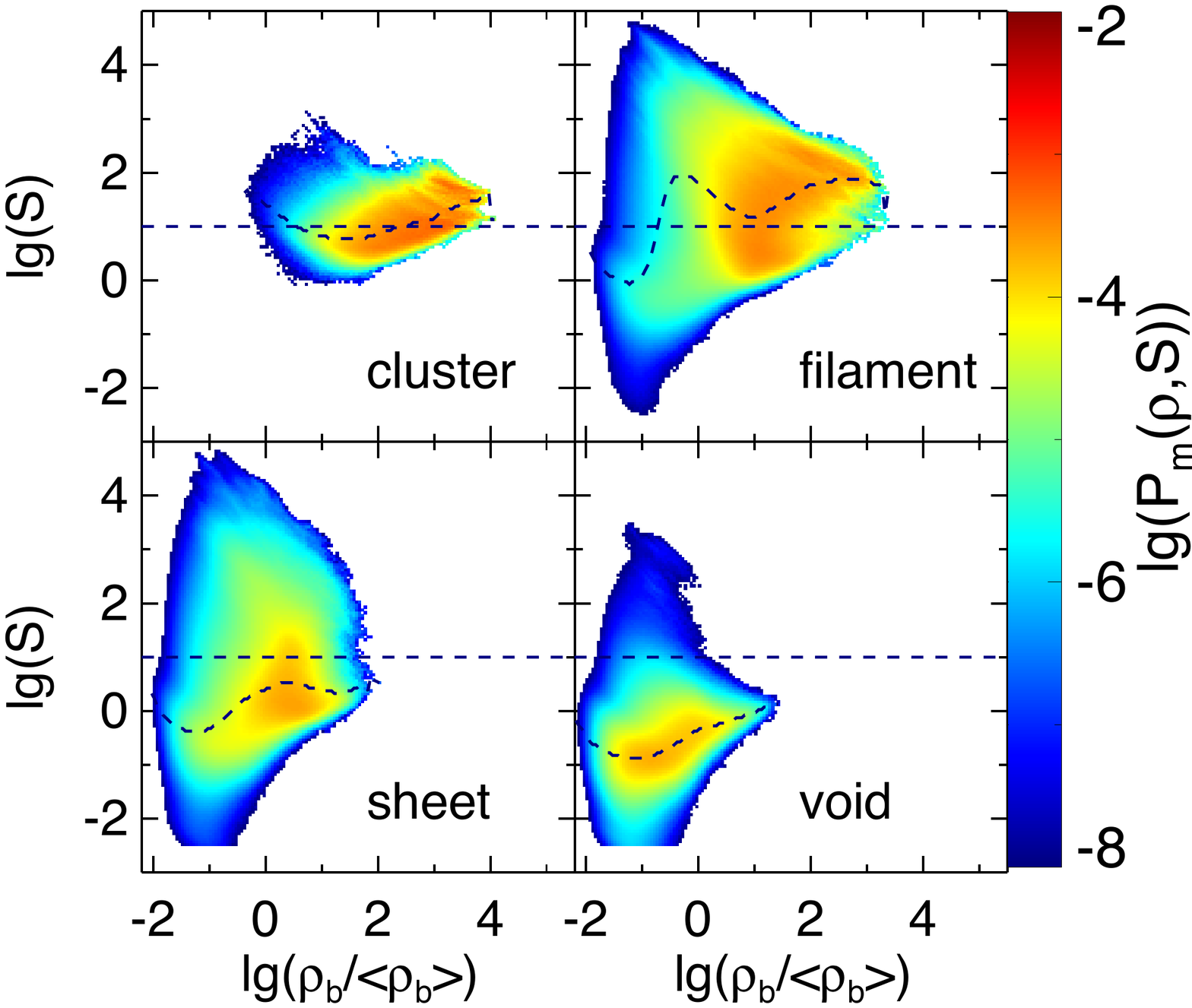}
\hspace{-3.0cm}
\includegraphics[width=0.40\textwidth]{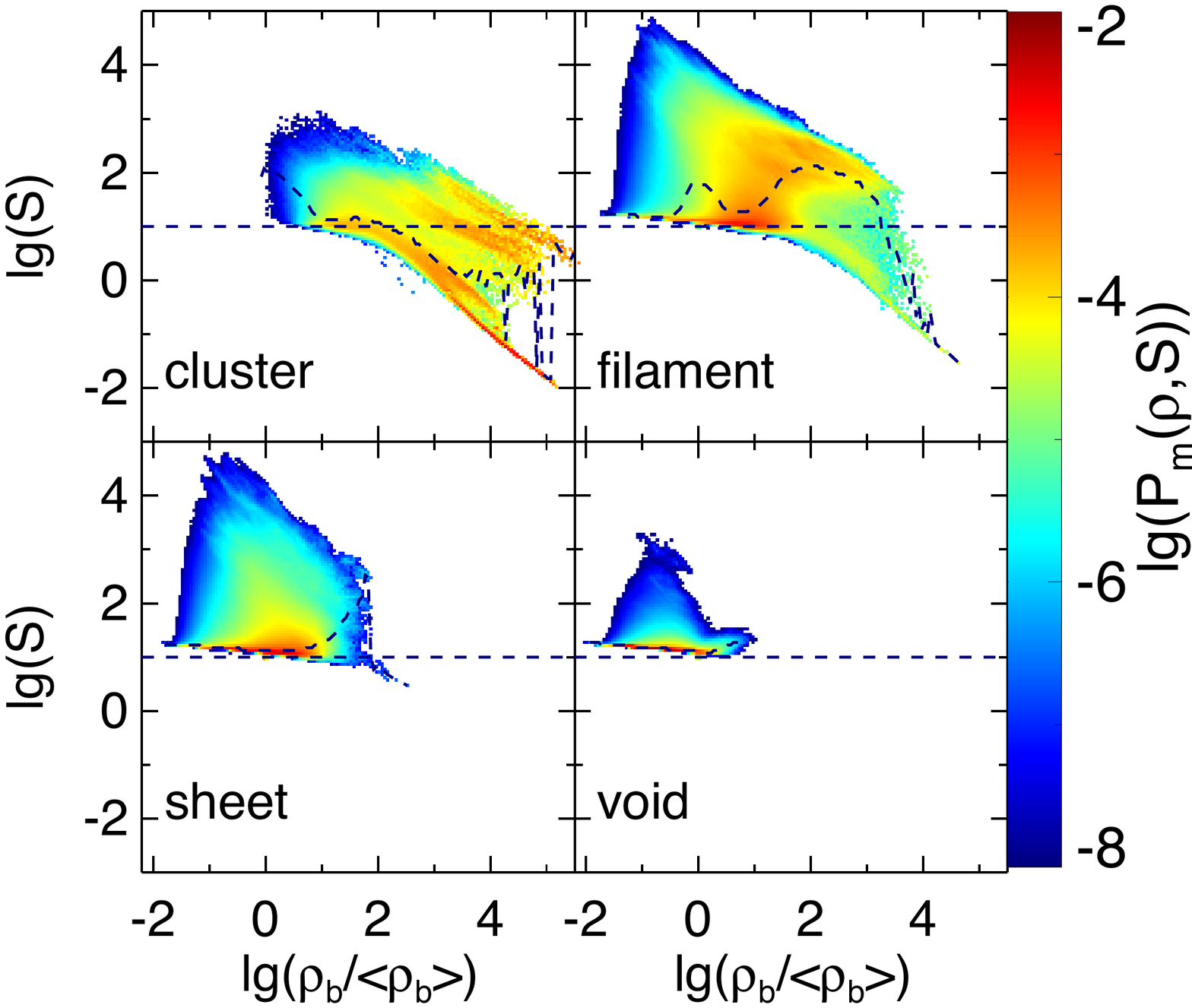}
}
\vspace{-2.0cm}
\caption{The mass weighted distribution in the density-entropy phase diagram of intergalactic medium residing in different environments in simulation L025-ada(Left) and L025-uvc(Right) at $z= 3.0, 2.0, 1.0, 0.0$(from top to bottom). Dashed curves indicate the median values of gas entropy in the corresponding density bin.}
\label{figure3}
\end{figure*}

Figure 2 further displays the phase diagrams of IGM in the density-temperature space for the four distinct components of the cosmic web respectively. The dashed curve in each plot gives the median values of gas temperature in the corresponding density bin. In L025-ada, the median temperature grows with the density increasing, following approximately a power law $T \propto \rho^{\alpha}$ with $\alpha \sim 1.2$ at $z \geq 2$. The mass fraction in filaments shows an evident growth since $z=2$. A rapid increase in temperature around $\rho=6-10$ at $z<2$ is also observed, which is in agreement with Valageas et al. (2002), and has made a primary contribution to the growth of WHIM gas. The associated gravitational heating due to filament formation should be the cause. 

The UVB in L025-uvc has heated up the diffuse gas effectively in the sheets and voids since $z \geq 3$ in comparison to L025-ada. The $T-\rho_b$ relation of diffuse IGM at high redshifts is dependent on the UVB model, although shock heating due to collapse would broaden the distribution. We fit the $T-\rho_b$ relation with a power law of $T(\rho_b)=T_0(\rho_b/\bar{\rho}_b)^{\gamma-1}$ in the range $10^{-0.5}<\rho_b/\bar{\rho}_b<10^{0.5}$ and $T<10^{4.3}\ $K. The values of $T_0$ and $\gamma$ are $1.54 \times 10^4\ $K and $1.55$ respectively at $z=2$. The index $\gamma$ is obviously compatible with the previous simulation works that had adopted the UVB of HM12(e.g., Lukic et al. 2015; Bolton et al. 2017). However, the $T_0$ in our simulation is moderately higher. This may be partially caused by the difference in fitted region, and the shock heating, especially in sheets and filaments. Nevertheless, this discrepancy would not change our conclusions significantly, as the preheating at high redshifts is ineffective in our simulations, which will be shown in the following sections.

The radiative cooling is efficient for the hot and dense gas with $\delta_b >100$ in both filaments and clusters in L025-uvc. The mass fractions of gas in various density-temperature phases in our simulations since $z=3$ are given in the Appendix. Our results are basically consistent with previous studies(e.g., Cen et al. 2006; Dave et al. 2010; more recently Snedden et al. 2016), but some differences can be observed at $z\leq0.5$. Since we are more concerned about the entropy of gas in different cosmic structures and their evolution, we will go straight to the statistics of entropy in the rest of this paper.

The left column of Figure 3 presents the density-entropy distributions in the four components of the cosmic web in L025-ada from $z=3.0$ to $0.0$. Visually, the median gas entropy is generally increasing from voids to sheets, and then to filaments, and lastly flattened into clusters. The last panel suggests that the gas with $S>10$\kevcm \, and $\delta_b<\delta_{th}$ are mostly residing in the filaments. A substantial part of those gas should be WHIM. A considerable fraction of the gas in the clusters also has $S>10$\kevcm, which should mainly exist in the form of hot halo gas, as judging from their density.  While in the sheets and voids, there is only a small amount of gas having $S>10$\kevcm.

Figure 3 indicates that the major increment of entropy due to gravitational heating occurs after $z=2$. Only a very small fraction of gas has an entropy higher than $10$\kevcm \, at $z>2$. While at $z=0$, a large amount of the gas in the filaments have builded a entropy $S>10$\kevcm \ . In the sheets and voids, however, the entropy of most of the gas still remains below $10$\kevcm \, in L025-ada. The entropy evolution of gas in different cosmic web components with a uniform UV background and radiative cooling can be found in the right column of Figure 3. The heating due to UVB has significantly raised the entropy of the gas with $\delta_b < \delta_{th}$ over $1$\kevcm \, since $z>3$, which happens in all the four components of the cosmic web. The cooling process is relatively more effective for the gas with $\delta_b > \delta_{th}$, which are mainly residing in filaments and clusters. The distribution of gas with entropy higher than $\sim 15-20$\kevcm \,, in the filaments in L025-uvc is mildly changed with respect to the adiabatic simulation. It can be easily understood as the thermal evolution of the gas in the filaments should mainly be governed by gravitational collapse. 

\begin{figure*}[htbp]
\vspace{-6.5cm}
\begin{center}
\includegraphics[width=0.80\textwidth]{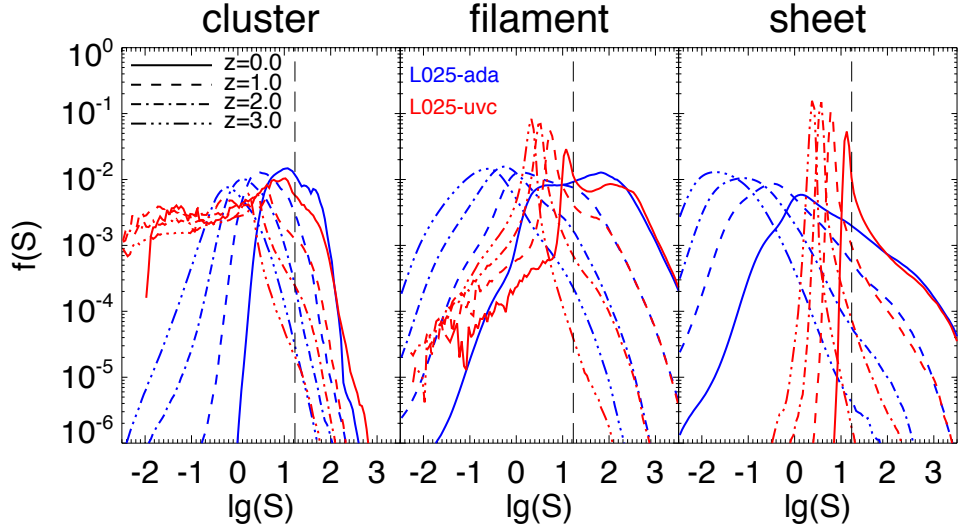}
\end{center}
\vspace{-6.0cm}
\caption{The mass weighted probability distributions of intergalactic medium as a function of entropy in L025-ada(Top), and L025-uvc(Bottom) at $z=3.0, 2.0, 1.0, 0.0$ in turn.}
\label{figure4}
\end{figure*}

\begin{figure*}[htbp]
\vspace{-3.5cm}
\begin{center}
\includegraphics[width=0.80\textwidth]{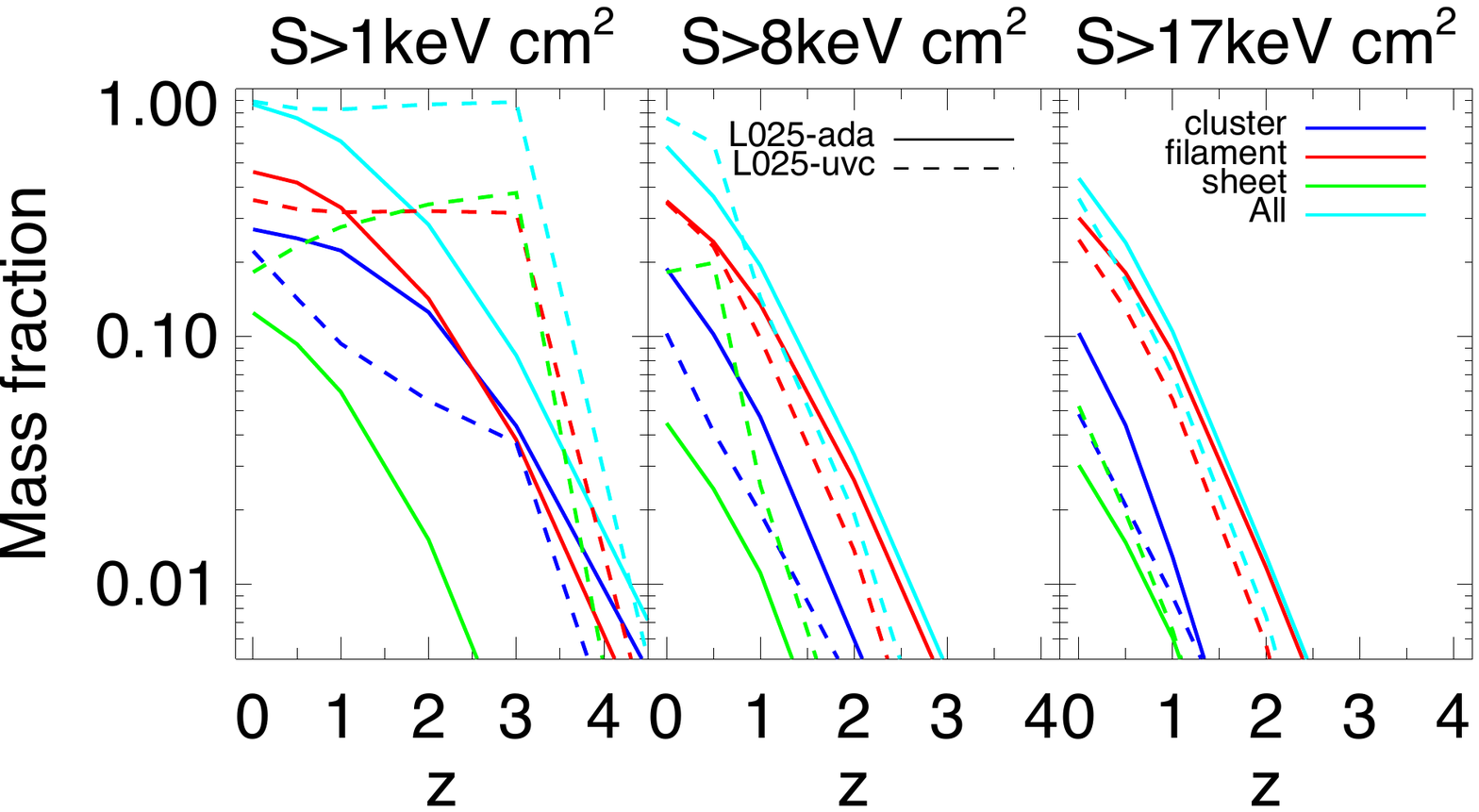}
\end{center}
\vspace{-9.5cm}
\caption{The mass fraction of intergalactic medium with entropy $S>1$\kevcm(Left), $S> 8$\kevcm(Middle), and $S> 17$\kevcm(Right) in simulation L025-ada(solid line) and L025-uvc(dashed line) since $z=5.0$. The total fraction is in the color of Cyan, and fractions contributed by gas in clusters, filaments and sheets are in the colors of blue, red, and green respectively.}
\label{figure5}
\end{figure*}

In Figure 4, we plot the mass weighted probability distributions of IGM as a function of entropy in our two simulations since $z=3$. As the mass fraction of gas contributed by the voids is rather small, only the distributions in the other three types of structures are plotted. When the gravitational collapse services as the sole heating mechanism, i.e., in L025-ada, the entropy of IGM shows extended distributions over a wide ranges for each cosmic web component. The peak value of gas entropy has been increasing gradually as the redshift decreasing. The probability distributions peak at about $1, 50, 10$\kevcm \, in sheets, filaments and clusters respectively at $z=0$. Plateaus are found in the probability distribution of gas entropy in filaments within the range $S \sim 1-100$\kevcm, and in clusters within $S \sim 3-30$\kevcm. The UVB generally heats up the gas with $\delta < \delta_{th}$ in all types of structures, leading to a jump in the pdf at 
$S \sim 10, 6, 5, 3$\kevcm at $z=0.0, 1.0, 2.0, 3.0$ correspondingly. Meanwhile, the radiative cooling makes a larger fraction of gas in the filaments and cluster having $S<1$\kevcm \, in comparison to L025-ada. The distribution above $S \sim 15-20$\kevcm \, is slightly modified by the UVB and cooling in sheets and filaments.  

To quantify the global effect of IGM heating due to gravitation and UV background, we investigate specifically the mass fraction of IGM with entropy $S>1$\kevcm, $S > 8$\kevcm, and $S> 17$\kevcm.  Lu \& Mo (2007) demonstrated that a preheating model with the circum-halo gas heated up to entropy $S=8$\kevcm \, would lead to the mass of cooled gas $M_{cool}$ scales with halo mass M as $M_{cool} \propto M^2$ in low-mass halos $M<10^{12} M_{\odot}$, and hence could help to explain the observed faint-end slope of galaxy luminosity function. The last threshold value, $17$\kevcm, is the constant entropy value on the right hand side of eqn(1), and is close to the predicted gas entropy due to the collapse of sheets at $z \lesssim 2$ in Mo et al. (2005), and is also the typical entropy required for halos with mass $M=10^{12} h^{-1} M_{\odot}$ in the preventative model given in Lu et al.(2015). Figure 5 gives the mass fraction of preheated IGM since $z=4.0$. In L025-ada at $z=0$, about $90\%, 60\%, 45\%$ of the IGM has an entropy larger than $1, 8, 17$\kevcm \, respectively. The filaments hosts about half of the IGM with $S>8$\kevcm, and about $70\%$ of those with $S>17$\kevcm. The corresponding shares contributed by clusters are $\sim 33\%, 20\%$ respectively for $S>8, 17$\kevcm. In other word, the high levels of IGM entropy produced by gravitational heating is contributed predominantly by the formation of filaments. 

The mass fractions of heated IGM drop rapidly as the redshift increasing. Around $20\%$ and $10\%$ of the IGM are heated up to $S>8, 17$\kevcm \, at $z=1.0$, while at redshift $z=2.0$ these fractions drop further to $3 \%$ and $1\%$ only. The dominant role of filaments is slightly more evident at high redshifts. The fractions of gas with $S>8, 17$\kevcm \, are somewhat altered by the UV background and cooling. It is actually in accordance with the distribution in density-entropy space as shown in Figure 3. The entropy of hot and dense gas in filaments and clusters are reduced by the radiative cooling. The diffuse gas with $\delta_b< 100$ in voids, sheets and even filaments are heated up by the UV photos. Consequently, the total mass fractions of IGM with $S>8, 1$\kevcm \, are enhanced since $z=0.5, 3.0$ respectively due to the UV heating.

In summary, about $60\%, 45\%$ of the IGM are heated up to $S>8, 17$\kevcm \, at $z=0$ due to gravitational collapsing. The fractions of IGM heated up to $S>8.0, 17.0$\kevcm \, drop rapidly while going to high redshifts, and are only a few of percents at $z=2$, lower than the prediction in Mo et al.(2005). Moreover, the formation of filaments should be the most effective gravitational preheating process, other than sheets. In addition, the ionizing heating from the uniform UV background (Haardt \& Madau 2012) has a minor effect on increasing the mass fraction of IGM with $S>17.0$\kevcm. 

\section{Preheating and cooling of circum-halo gas }
In this section, we further explore the preheating of halo and circum-halo gas, and estimate the impact on cooling, which is crucial for understanding the gas accretion of dark matter halos, and consequently the star formation in galaxies. 

\begin{figure*}[htbp]
\vspace{-1.5cm}
\gridline{
\hspace{-1.0cm}
\includegraphics[width=0.55\textwidth]{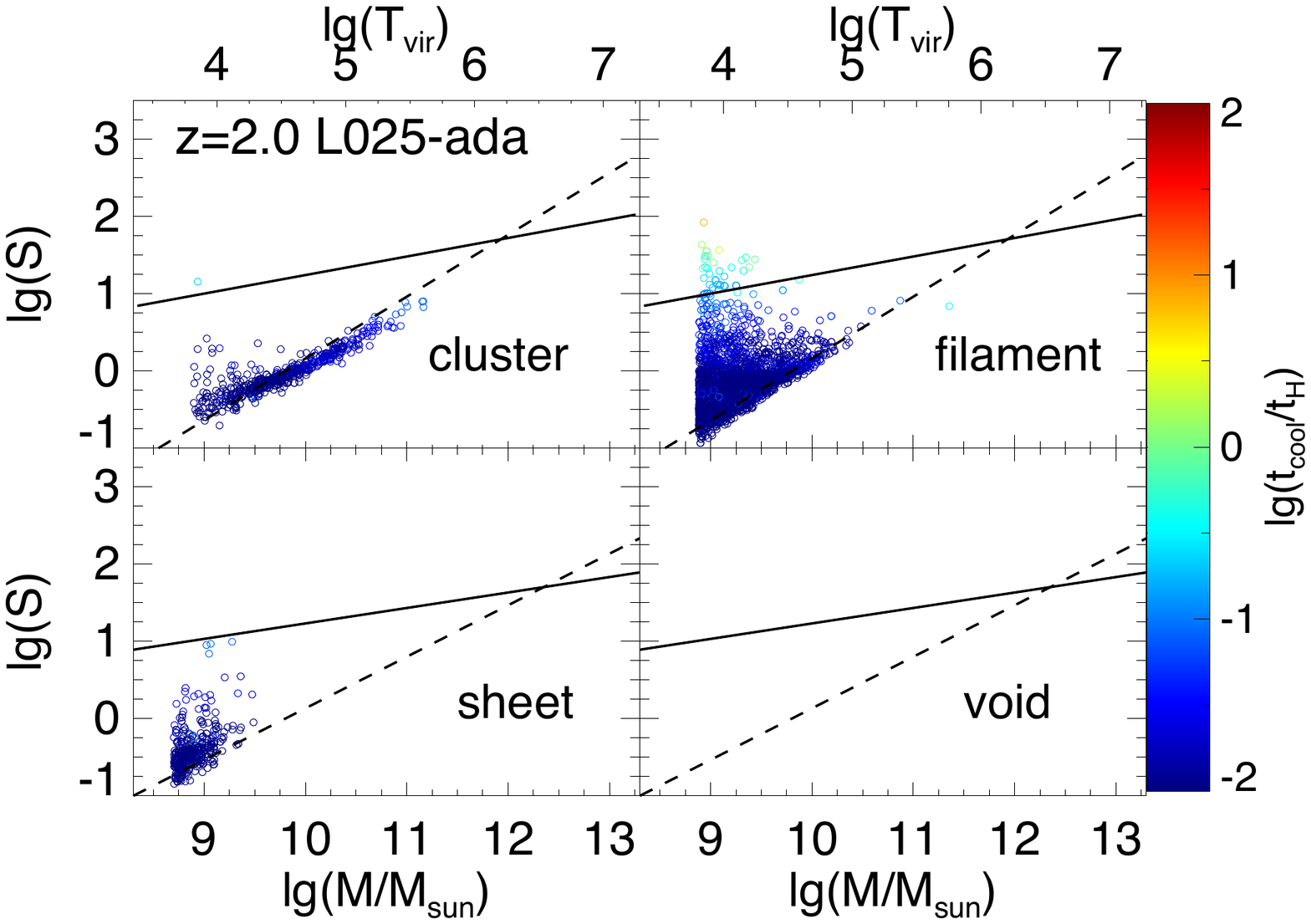}
\hspace{-1.0cm}
\includegraphics[width=0.55\textwidth]{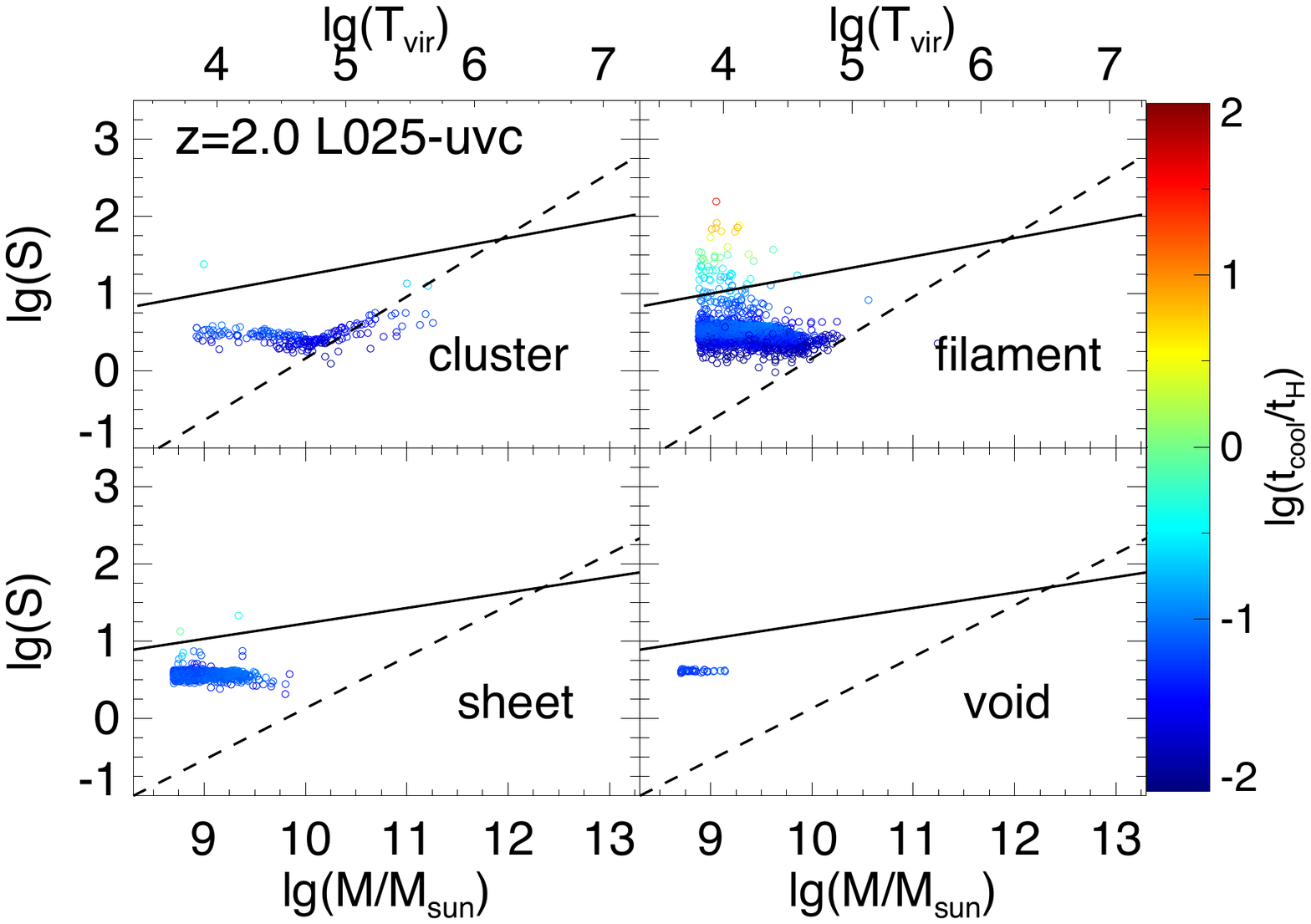}
}
\vspace{-7.0cm}
\gridline{
\hspace{-1.0cm}
\includegraphics[width=0.55\textwidth]{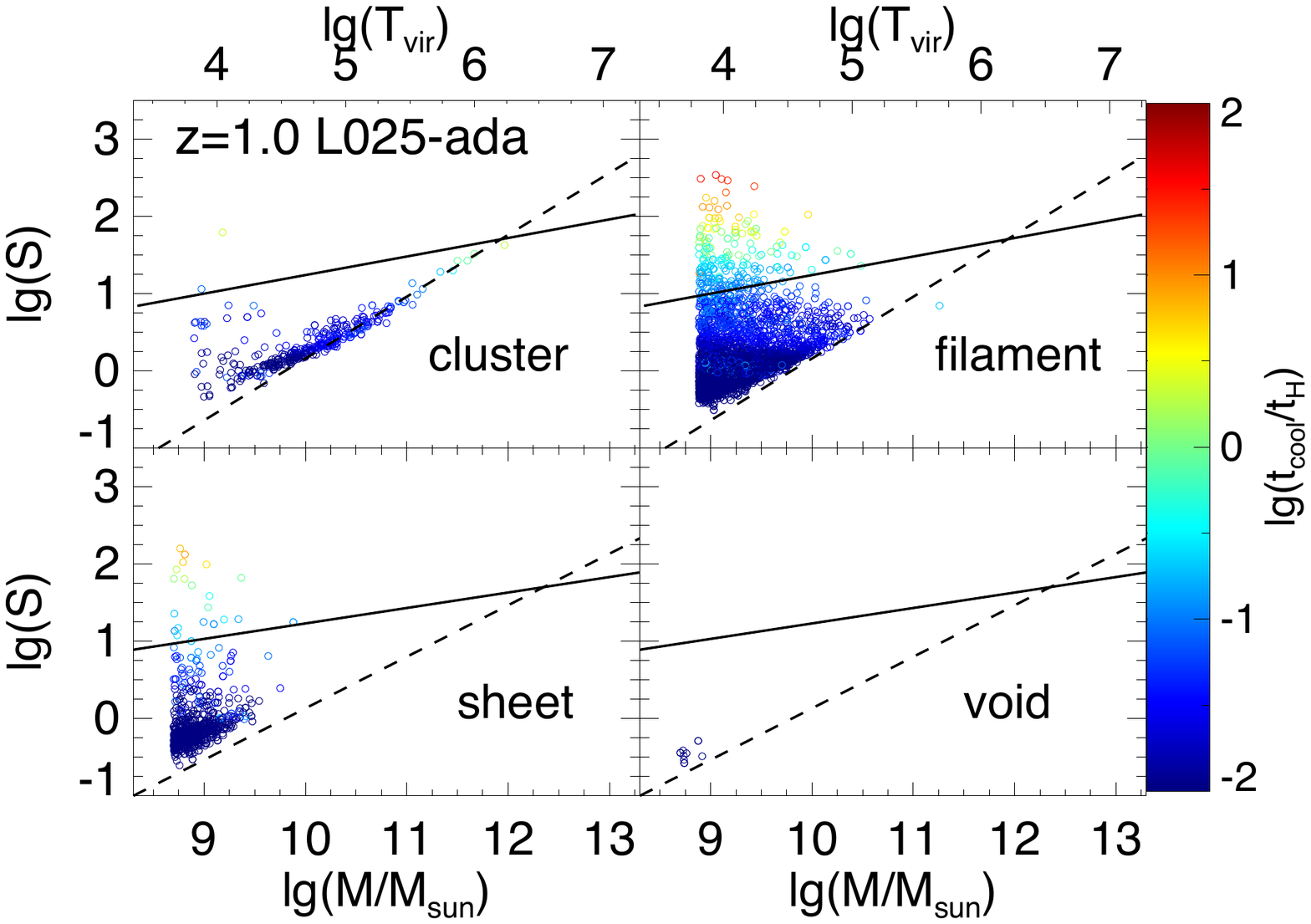}
\hspace{-1.0cm}
\includegraphics[width=0.55\textwidth]{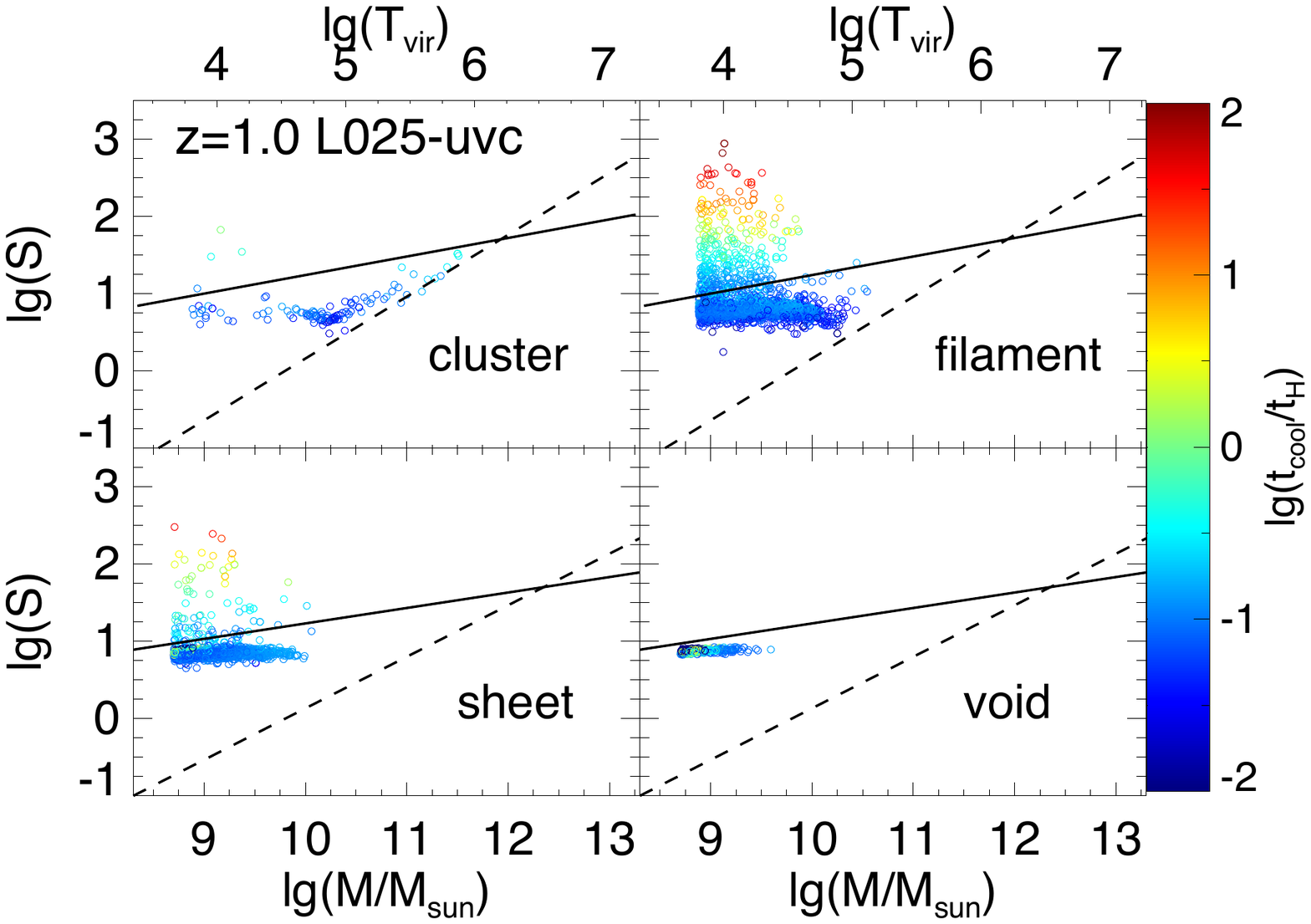}
}
\vspace{-7.0cm}
\gridline{
\hspace{-1.0cm}
\includegraphics[width=0.55\textwidth]{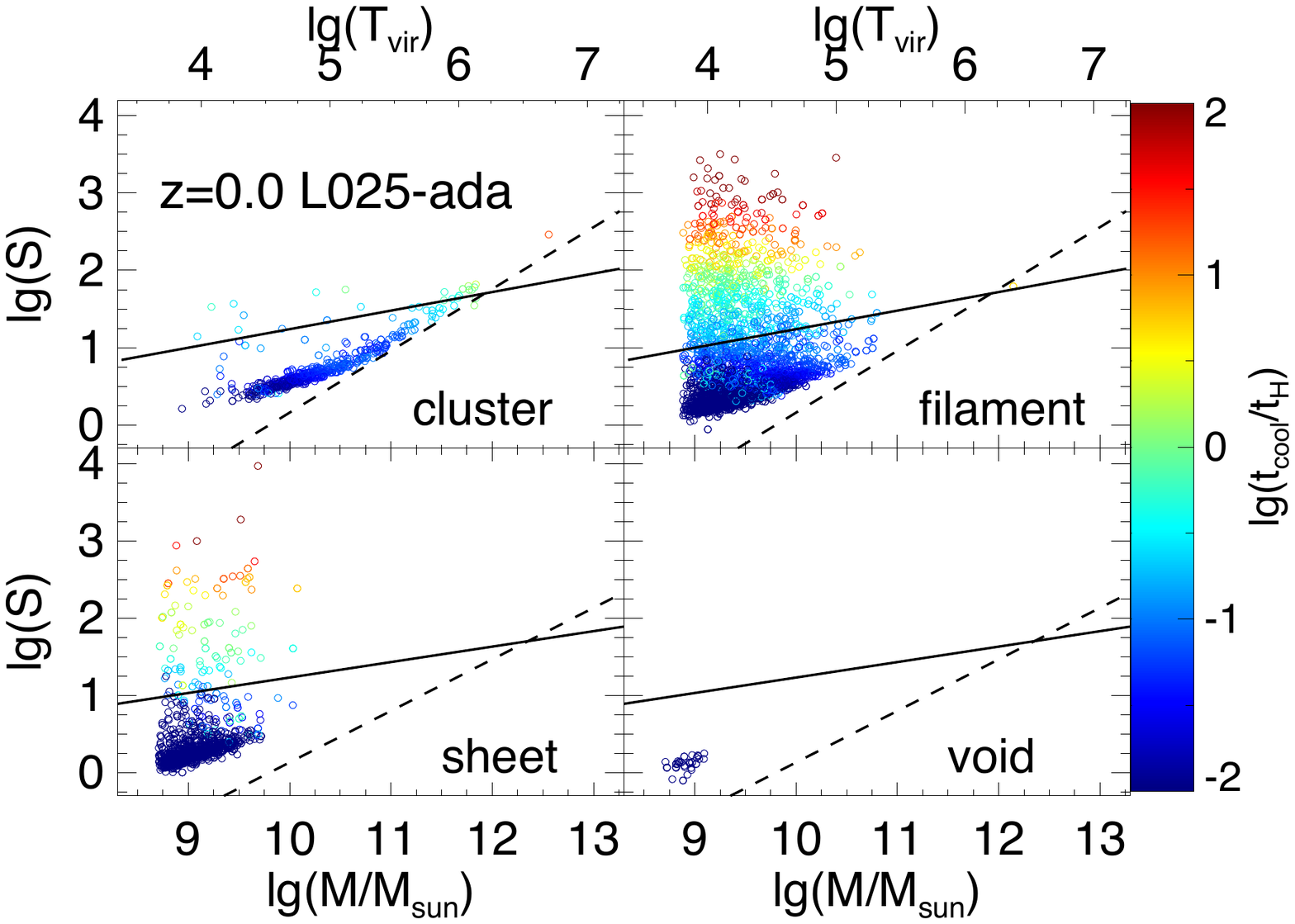}
\hspace{-1.0cm}
\includegraphics[width=0.55\textwidth]{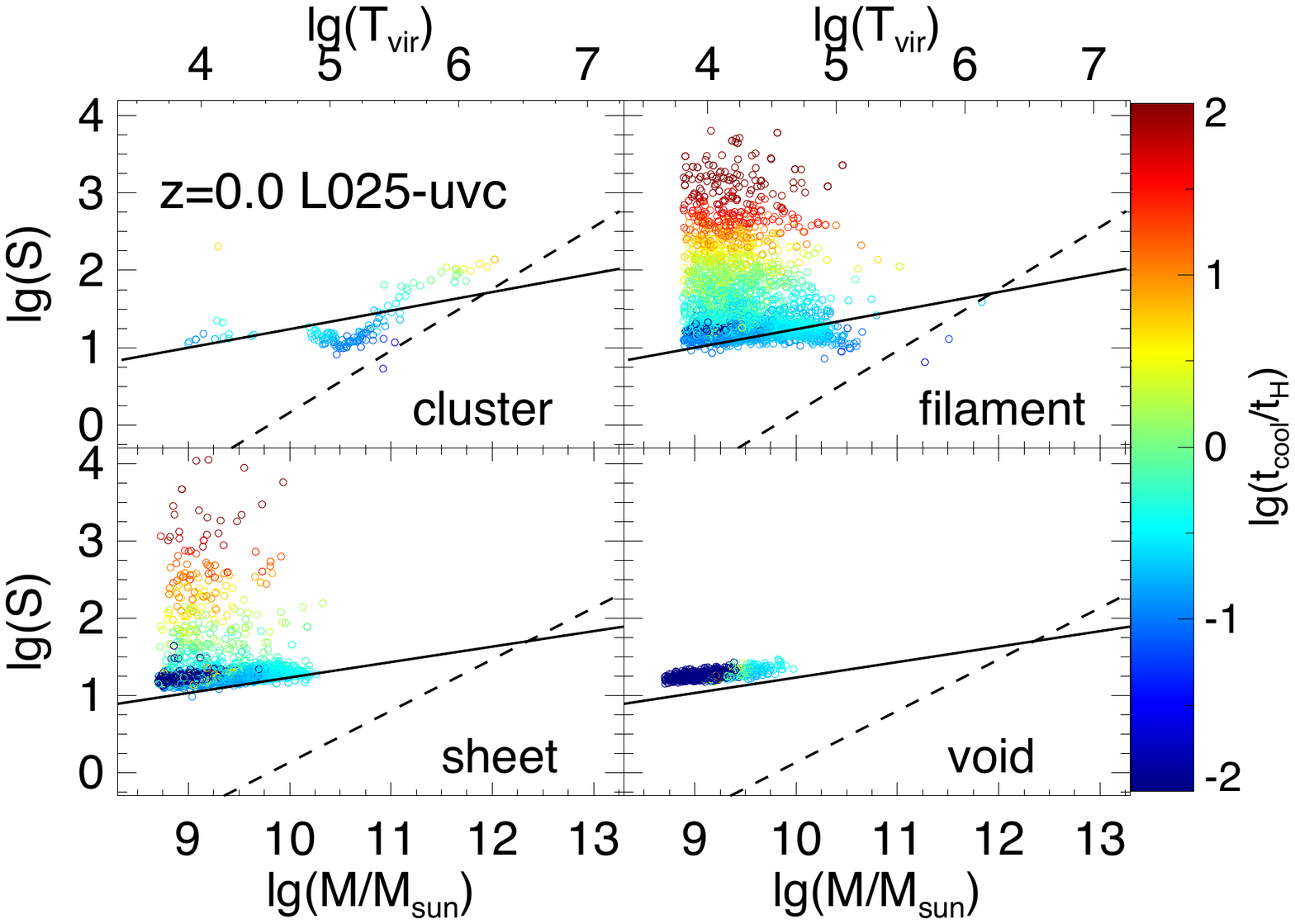}
}
\vspace{-6.0cm}
\caption{The mass weighted average entropy $S_{\rm{cir}}$ of gas within $R_{vir}<r<2R_{vir}$ of 5000 randomly selected dark matter halos, assigned to different cosmic environments, at $z=2.0, 1.0, 0.0$ in L025-ada(Left) and L025-uvc. Dashed lines indicate the virial entropy, and the solid lines indicate the value given by eqn(2). Halos are color-coded by the ratio of cooling time to the Hubble time.}
\label{figure6}
\end{figure*}

\begin{figure*}[htbp]
\vspace{-1.5cm}
\gridline{
\hspace{-1.0cm}
\includegraphics[width=0.55\textwidth]{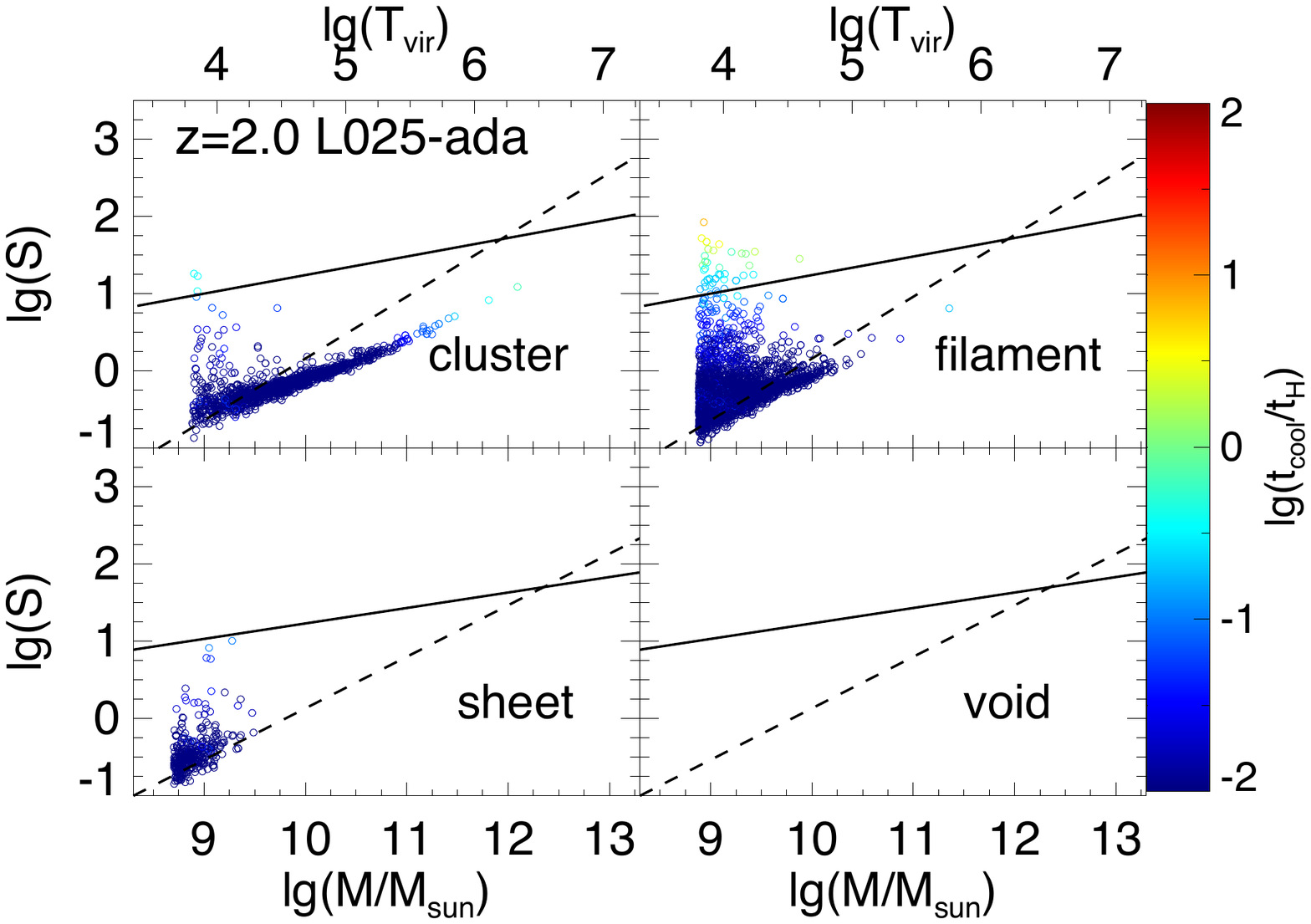}
\hspace{-1.0cm}
\includegraphics[width=0.55\textwidth]{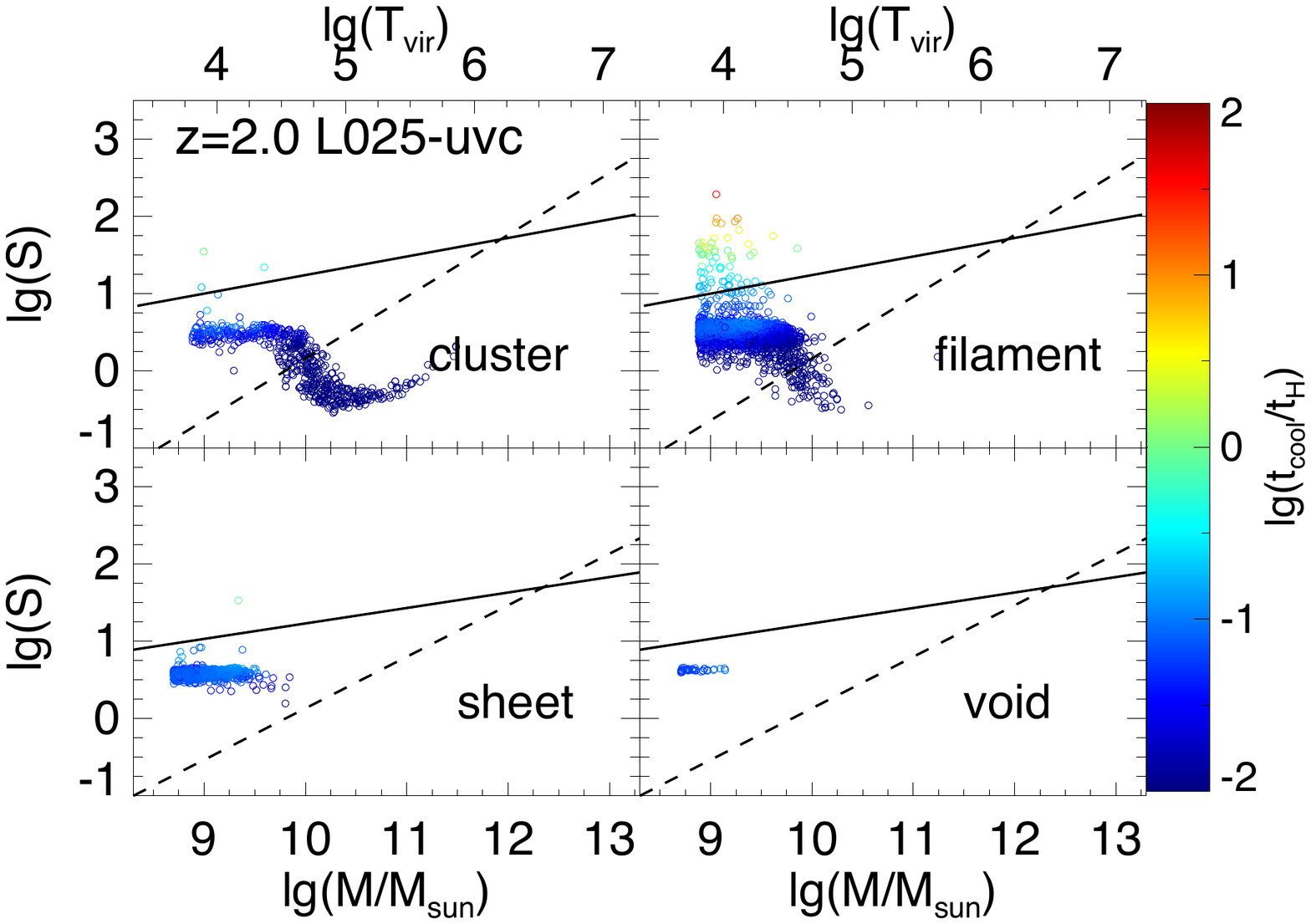}
}
\vspace{-7.0cm}
\gridline{
\hspace{-1.0cm}
\includegraphics[width=0.55\textwidth]{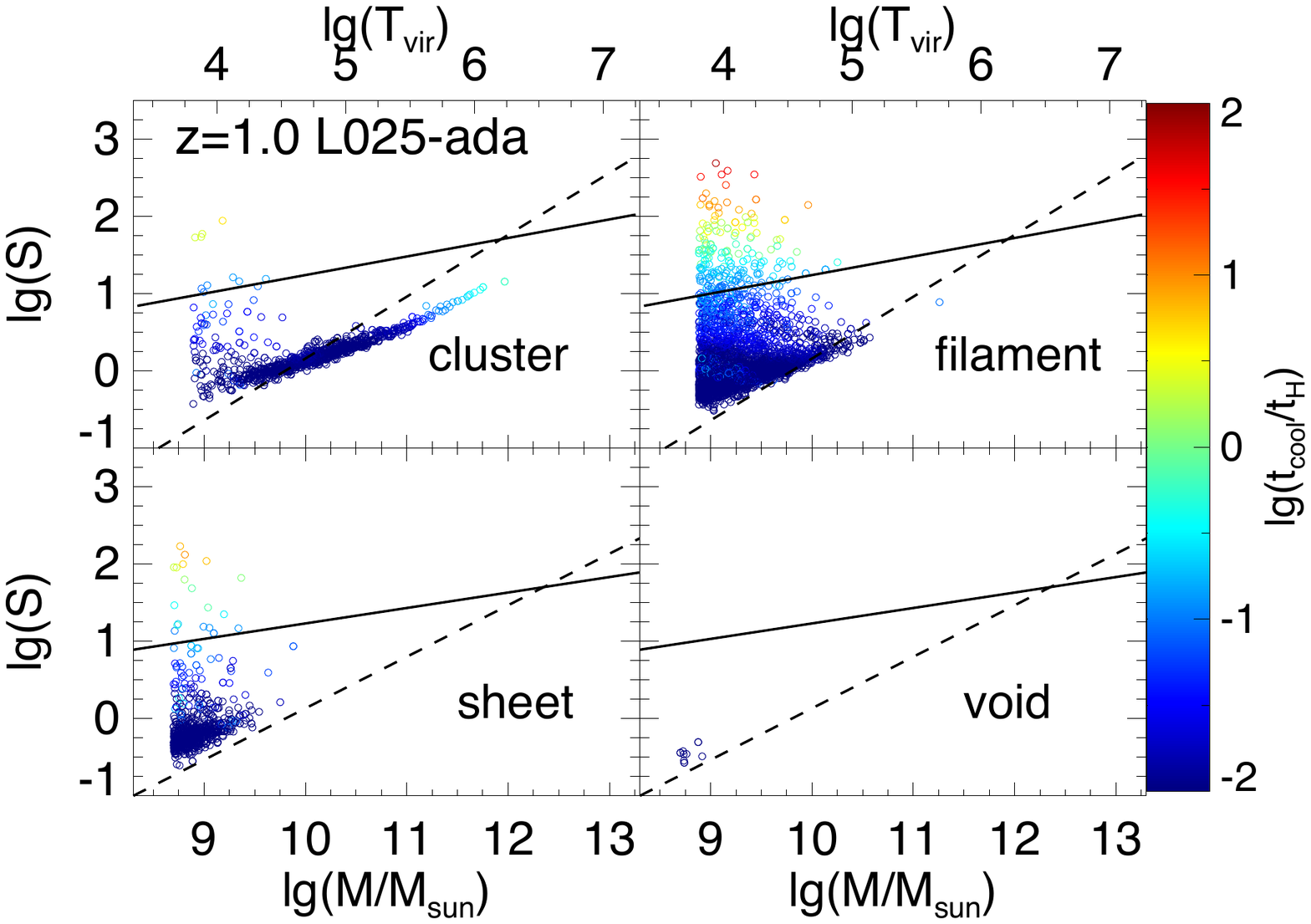}
\hspace{-1.0cm}
\includegraphics[width=0.55\textwidth]{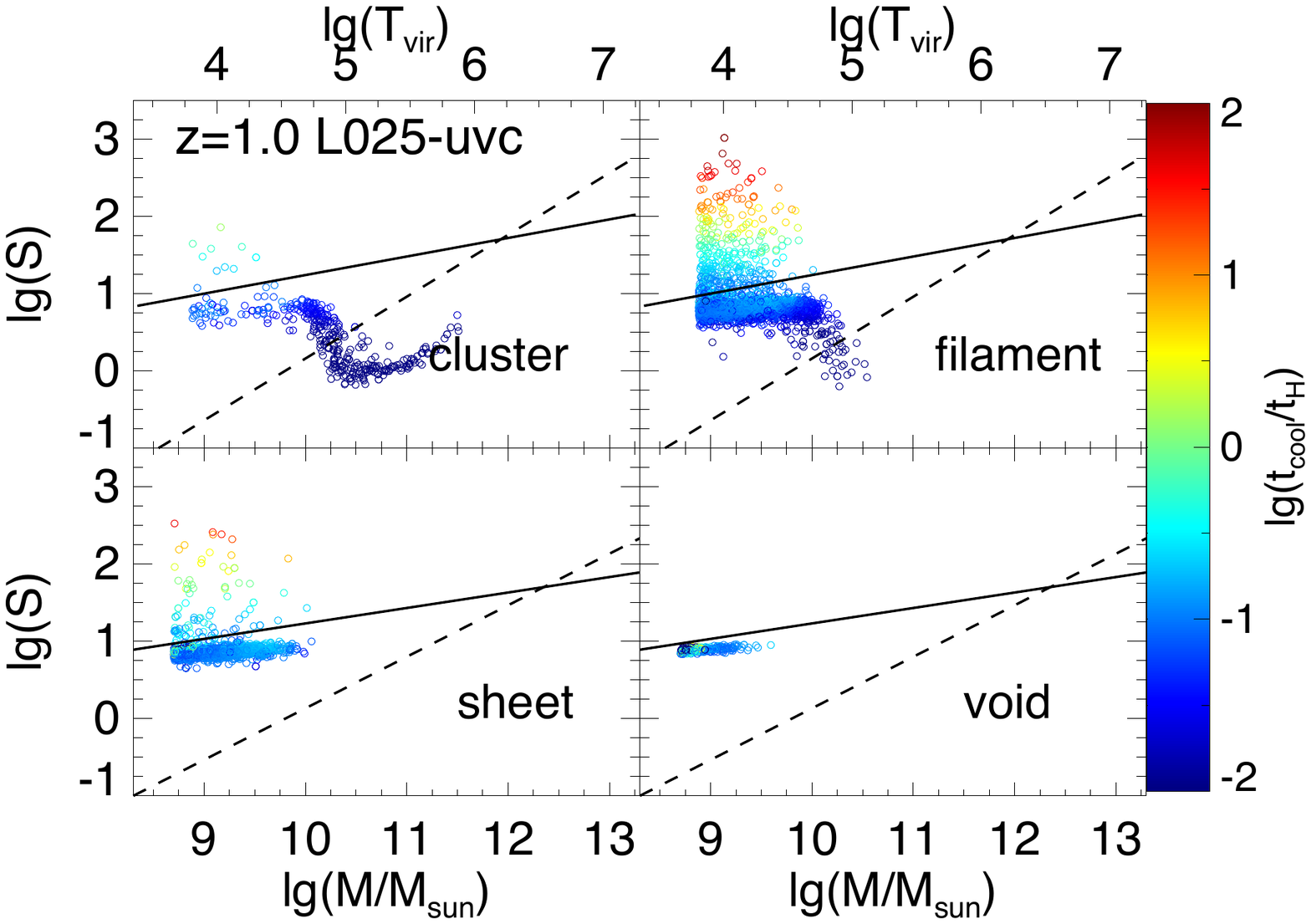}
}
\vspace{-7.0cm}
\gridline{
\hspace{-1.0cm}
\includegraphics[width=0.55\textwidth]{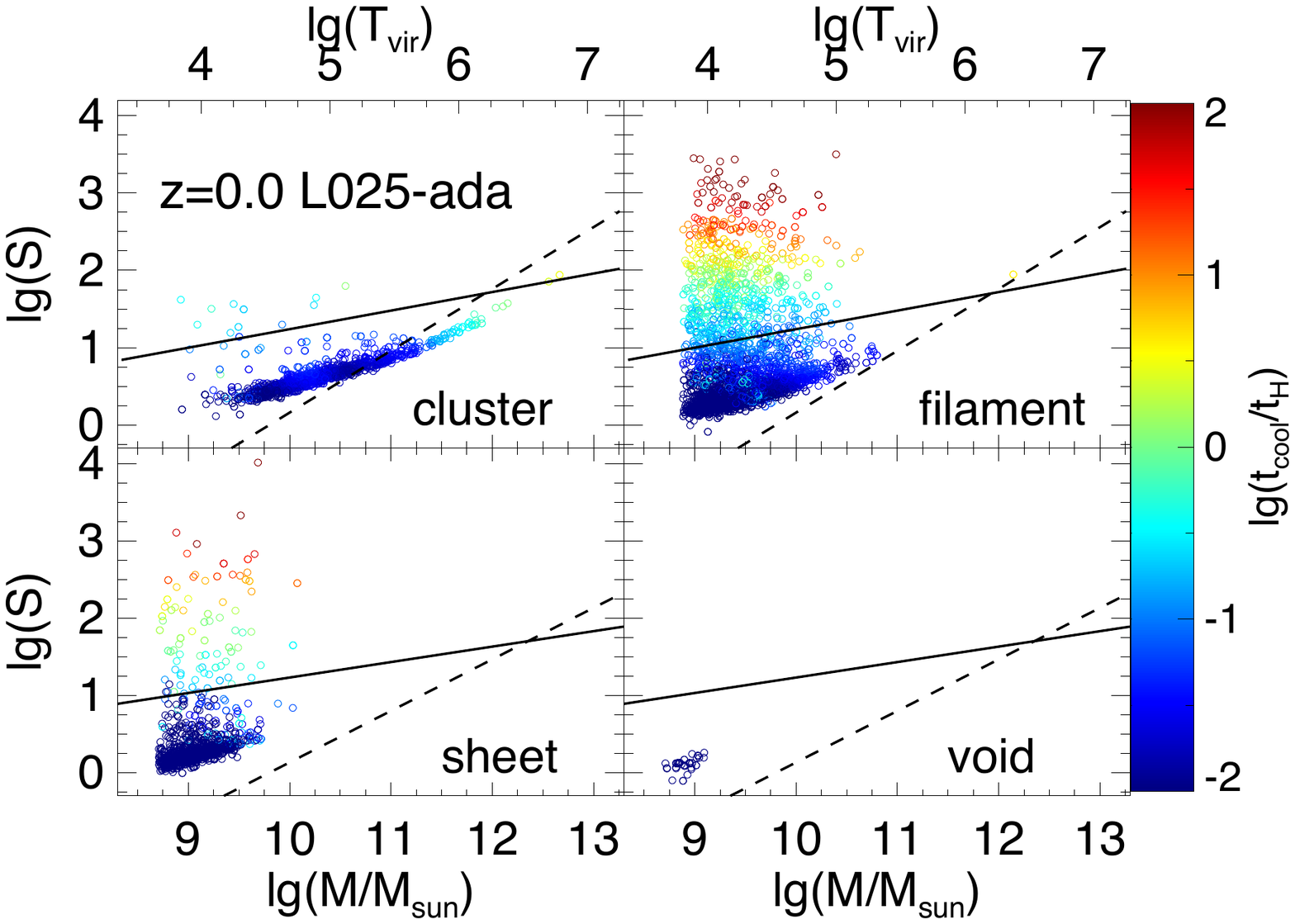}
\hspace{-1.0cm}
\includegraphics[width=0.55\textwidth]{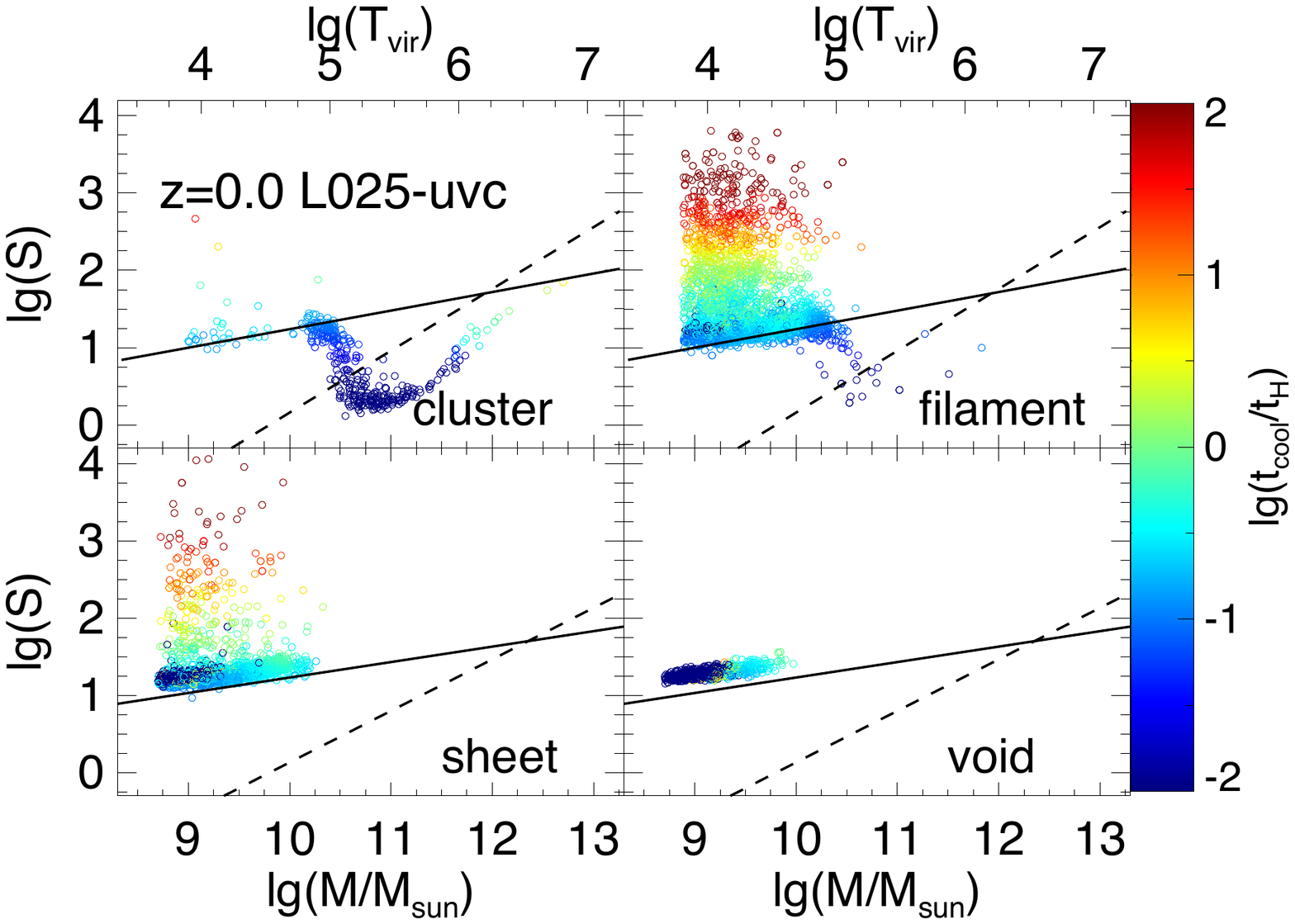}
}
\vspace{-6.0cm}
\caption{Same as Figure 6, but for the mass weighted average entropy, $S_h$ , of gas within the virial radius of dark matter halos.}
\label{figure7}
\end{figure*}

\subsection{Entropy of halo and circum-halo gas in different cosmic web environments}

To make a comparison with the preventative feedback models and previous simulations(e.g., Cen 2011), the entropy of halo gas $S_{h}$ and circum-halo gas $S_{\rm{cir}}$ are used as indicators. More specifically, we calculate the average density and mass weighted average temperature of gas residing within each dark matter halo, and then obtain $S_{h}$ according to eqn.(1). Similarly, the average density and temperature of the IGM in the shell with $R_{vir}<r<2R_{vir}$ that surrounding each halo are calculated to obtain $S_{\rm{cir}}$. The latter characterizes the thermal state of gas right before collapsing into halos, and hence can serve as a good indicator of preheating. To lower the effect of the limited resolutions in our fixed-grid simulations, we probe only those halos that are composed of more than $250$ particles. Figure 6 shows the entropy of circum-halo gas, $S_{\rm{cir}}$, of randomly selected 5000 dark matter halos in our two simulations at $z=2.0, 1.0$ and $0.0$. The circles are color-coded by the ratio of cooling time to the Hubble time, which will be discussed in detail in the next subsection. Only a few halos are more massive than $10^{12} M_\odot$ in our simulations, due to the limited box size. Most of the dark matter halos have been residing in filaments since $z=2$, which is consistent with previous works based on N-body simulations(e.g., Catun et al. 2014). The dashed lines in Figure 6 display the virial entropy, $S_{\rm{vir}}=T_{vir}/n_{vir}^{2/3}$, where $T_{vir}$ is the virial temperature and $n_{vir}$ is the mean gas density within a dark matter halo by assuming its baryon fraction taking the value of the cosmic mean $\Omega_{b}/\Omega_{m}$. The virial entropy indicates the heating produced by the collapse of halos themselves. 

To evaluate the strength of preheating in our simulations, we also plot another threshold entropy of circum-halo gas with solid lines given by
\begin{equation}
S_{\rm{pr}}=17(\frac{M_{\rm{vir}}}{10^{12}M_{\odot}})^{0.2} \frac{1}{1+(z/1.2)^2} \kevcm.
\end{equation}
where $M_{\rm{vir}}$ is the mass of dark matter halo. A formula like this was firstly introduced by Lu et al(2015) to account for various preheating processes in their preventative model of disc galaxies formation. These parameters had been chosen such that their semi-analytical model with preheated gas can be capable of reproducing the observed scaling relations of disc galaxies remarkably well. For the sake of simplicity, we use the mass of a halo at particular redshifts, instead of its final mass at $z=0$, to calculate the $S_{\rm{pr}}$ at $z$. This modification might lead to a deviation about $1$\kevcm \, from the original formula in Lu et al.(2015).  In L025-ada, the average entropy of circum-halo gas, $S_{\rm{cir}}$, of most halos are generally above the virial entropy $S_{\rm{vir}}$ at $z=2$. The excess with respect to $S_{\rm{vir}}$ could be attributed to the collapse of large scale structures, i.e., the formation of filaments and sheets. The preheating of circum-halo gas is relatively more significant for halos in filaments than those in sheets, clusters and voids, in agreement with the density-entropy distribution in different environments. The margin of $S_{\rm{cir}}$ over virial entropy increases gradually along time for $M_{\rm{vir}}<10^{11.0} M_{\odot}$, compatible with the growth scenario of IGM entropy toward low redshifts reported in the last section. However, only a few of halos are surrounded by gas that has been heated up to $S_{\rm{cir}}>S_{\rm{pr}}$ in L025-ada. 

The inclusion of UV background in L025-uvc have effectively improved the fractions of halos with $S_{\rm{cir}}>S_{\rm{pr}}$ for $M_{\rm{vir}}<10^{10.5} M_{\odot}$, especially in filaments, sheets and voids. This enhancement should be caused by the heating of gas with $\delta_b <100$ by the UVB since high redshifts, as shown in Figure 3. Gas that was heated by UV at early time will go through further gravitational heating while accreting into more over dense region at low redshift. In L025-uvc, a shallow trough in $S_{\rm{cir}}$ is found for halos with mass $10^{10.5}<M_{\rm{vir}}<10^{11.0} M_{\odot}$ in clusters, which should result from the highly efficient radiative cooling. Actually, the radiative cooling function peaks at around $2\times 10^{4}, 10^{5}$ K(e.g., see Sutherland \& Dopita 1993). In the presence of UVB, the temperature of most of the IGM can cross over the first peak, i.e., $2\times 10^4\ $K. However, UV heating can hardly lift the fraction of IGM that can reach the second peak, $10^{5}\ $K, close to the virial temperature of a halo with mass $10^{10.5} M_{\odot}$. A shallower trough of $S_{\rm{cir}}$ with respect to $S_{\rm{pr}}$ can be also observed for halos with masses of $M_{\rm{vir}}>10^{10.0} M_{\odot}$ in the filaments. 

Figure 7 shows the mass weighted average entropy $S_{h}$ of gas within the dark matter halos. There is only minor differences between $S_{h}$ and $S_{\rm{cir}}$ in the adiabatic simulation. Once radiative cooling is on, $S_{h}$ shows distinct drop for halos with masses greater than a couple of $10^{10} M_{\odot}$. The drop is more significant in the environment of clusters. For these halos, the baryonic matter would have experienced distinct radiative cooling within the virial radius, and hence the entropy drops away from $S_{\rm{cir}}$ dramatically. Once again, this trend matches with the mass distribution in the density-temperature and density-entropy plane shown in section 3. The cooling is very effective for gas with $\delta \gtrsim 100$ and $T \sim 10^{4.5}-10^{5.0}$ K, which becomes more violent with the increasing over-density. 

\begin{figure*}[htbp]
\vspace{-8.8cm}
\begin{center}
\includegraphics[width=0.95\textwidth]{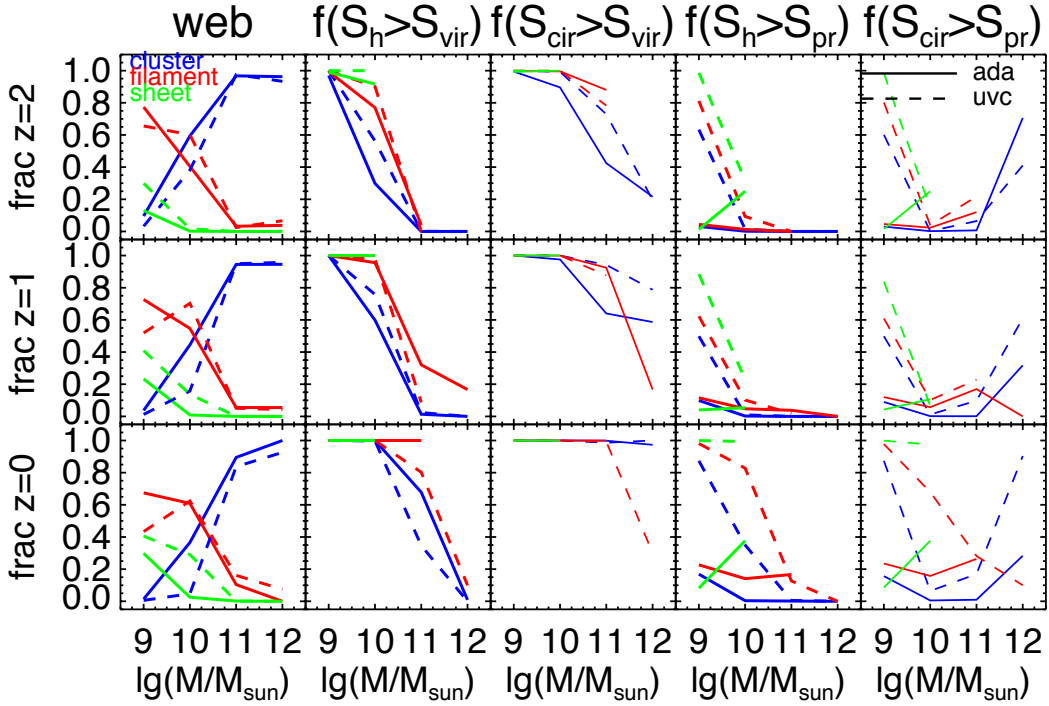}
\end{center}
\vspace{-3.5cm}
\caption{Left column: The fraction distribution of halos in different structures at $z=2.0, 1.0, 0.0$(from top to bottom). Blue, red and green colors indicates fractions in the environments of clusters, filaments and sheets respectively. Middle two columns: The fraction of halos that the mass weighted average entropy of gas within $r<R_{vir}$, and $R_{vir}<r<2R_{vir}$ are higher than the virial entropy in mass bins $10^{8.5}-10^{9.5} M_{\odot}$, $10^{9.5}-10^{10.5} M_{\odot}$, $10^{10.5}-10^{11.5} M_{\odot}$, and $10^{11.5}-10^{12.0} M_{\odot}$. Right two columns: Same as the Middle, but for gas entropy that are higher than the vaule $S_{\rm{pr}}$ given by eqn(2). }
\vspace{-0.2cm}
\label{figure8}
\end{figure*}

We plot the fractions of halos with $S_{\rm{cir}}$, and $S_{h}$ higher than $S_{\rm{vir}}$, and $S_{\rm{pr}}$ respectively at redshift $2, 1, 0$ in Figure 8. The halos are assigned into four mass bins, $10^{8.5}-10^{9.5} M_{\odot}$, $10^{9.5}-10^{10.5} M_{\odot}$, $10^{10.5}-10^{11.5} M_{\odot}$, and $10^{11.5}-10^{12.5} M_{\odot}$. The left column shows the halo distributions in different components of the cosmic web . For halos less massive than $\sim 10^{10.5} M_\odot$, the primary hosting structure is filaments. Massive halos tend to residing in clusters. The similar trend has been reported in the literature(e.g., Cautun et al. 2014), but the break halo mass is much higher in their work. The relative smaller box size of our simulations results in smaller collapsed structures, including halos and filaments.  Comparing to in L025-ada,  more halos are found in filaments and sheets in L025-uvc. The total number of halos in the last mass bin is small, and so does for halos in sheets those are more massive than $\sim 10^{10.0} M_\odot$. Therefore, statistics including fractions would suffer from notable random noise in corresponding bins. 

The 2nd and 3rd columns of Figure 8 show the fractions of $S_{h}>S_{\rm{vir}}$, and $S_{\rm{cir}}>S_{\rm{vir}}$ in the environments of clusters, filaments and sheets. At $z=2$, more than $80 \%$ of the halos with $M<10^{10.5} M_{\odot}$ have a $S_{\rm{cir}}$ higher than the virial entropy, and the fraction is lower for more massive halos, falling to $<30 \%$ for $10^{11.5}-10^{12.5} M_{\odot}$. The fractions grow with decreasing redshifts, and are nearly $100 \%$ at $z=0$ for all the halos. On the other hand, the fraction of $S_{h}>S_{\rm{vir}}$ is lower than that of $S_{\rm{cir}}>S_{\rm{vir}}$, about $\sim 75\%$ and $\sim 0\%$ for halos with $10^{10.5}-10^{11.5} M_{\odot}$, and $10^{11.5}-10^{12.5} M_{\odot}$ respectively. In L025-uvc, cooling process may have led the fraction of $S_{h}>S_{\rm{vir}}$ to decrease by $\sim 30\%$ for halos with $10^{10.5}-10^{11.5} M_{\odot}$ at $z=0$. 

The fractions of halos with entropy above $S_{\rm{pr}}$ are given in the right two columns of Figure 8. When the gravitational collapse is the sole preheating mechanism, the fractions with $S_{\rm{cir}}>S_{\rm{pr}}$, and $S_{h}>S_{\rm{pr}}$ are about $\sim 15-20 \%$ for halos with $M_{\rm{vir}}<10^{10.5} M_{\odot}$ at $z=0$, and drop rapidly toward high redshifts. A trough appears in the halo mass range of $10^{10.5}-10^{11.5} M_{\odot}$. Some of the most massive halos in our simulations, $M_{vir}\sim 10^{12.5} M_{\odot}$, are also surrounded by circum-halo gas with entropy higher than $S_{\rm{pr}}$. But for these halos, the entropy of gas surrounding halos is close to the virial entropy, which is higher than $S_{\rm{pr}}$. Namely, the heating due to the formation of halos themselves should be dominant over the preheating due to the collapse of filaments and sheets hosing those halos. The UV background heating significantly increase the fraction of halos with $S_{\rm{cir}}>S_{\rm{pr}}$, and $S_{h}>S_{\rm{pr}}$ to nearly $70 \%$ for $M_{\rm{vir}}<10^{10.5} M_{\odot}$ at $z=0$, and to $50 \%$ for $M_{\rm{vir}}<10^{9.5} M_{\odot}$ at $z>0$.

\subsection{Cooling of halo and circum-halo gas}

\begin{figure*}[htbp]
\vspace{-2.0cm}
\hspace{-0.0cm}
\includegraphics[width=0.55\textwidth]{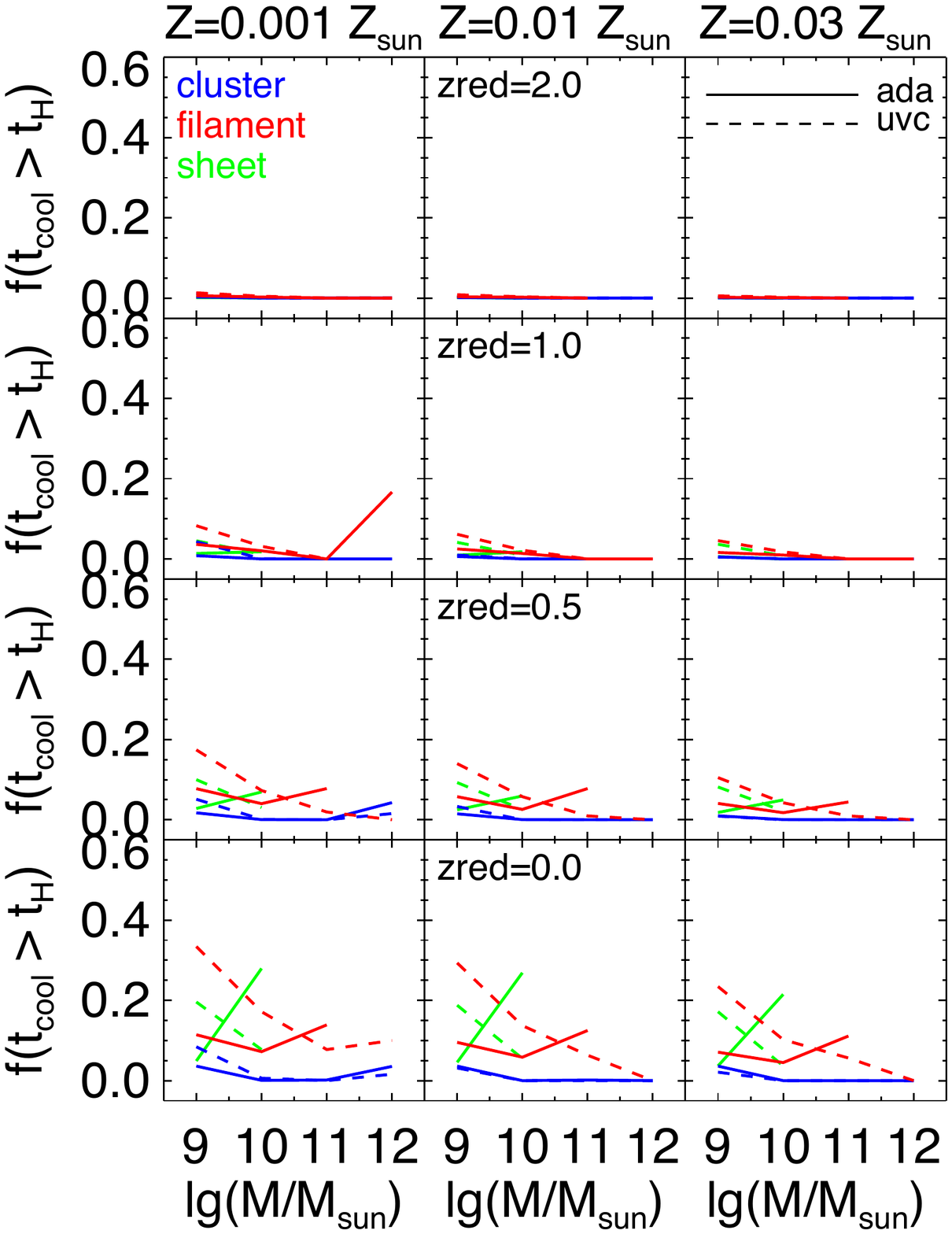}
\hspace{-1.4cm}
\includegraphics[width=0.55\textwidth]{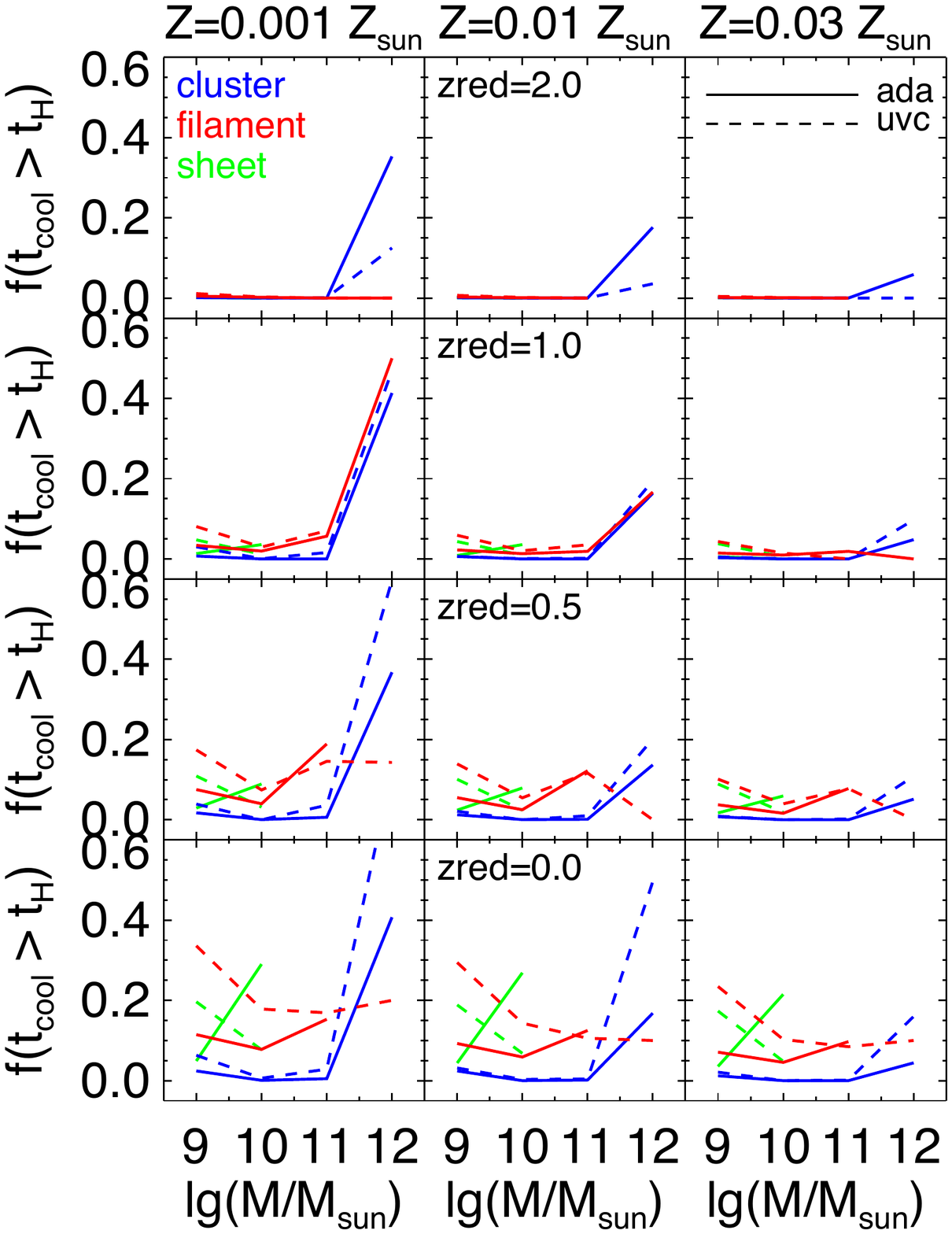}
\vspace{-2.0cm}
\begin{center}
\includegraphics[width=0.55\textwidth]{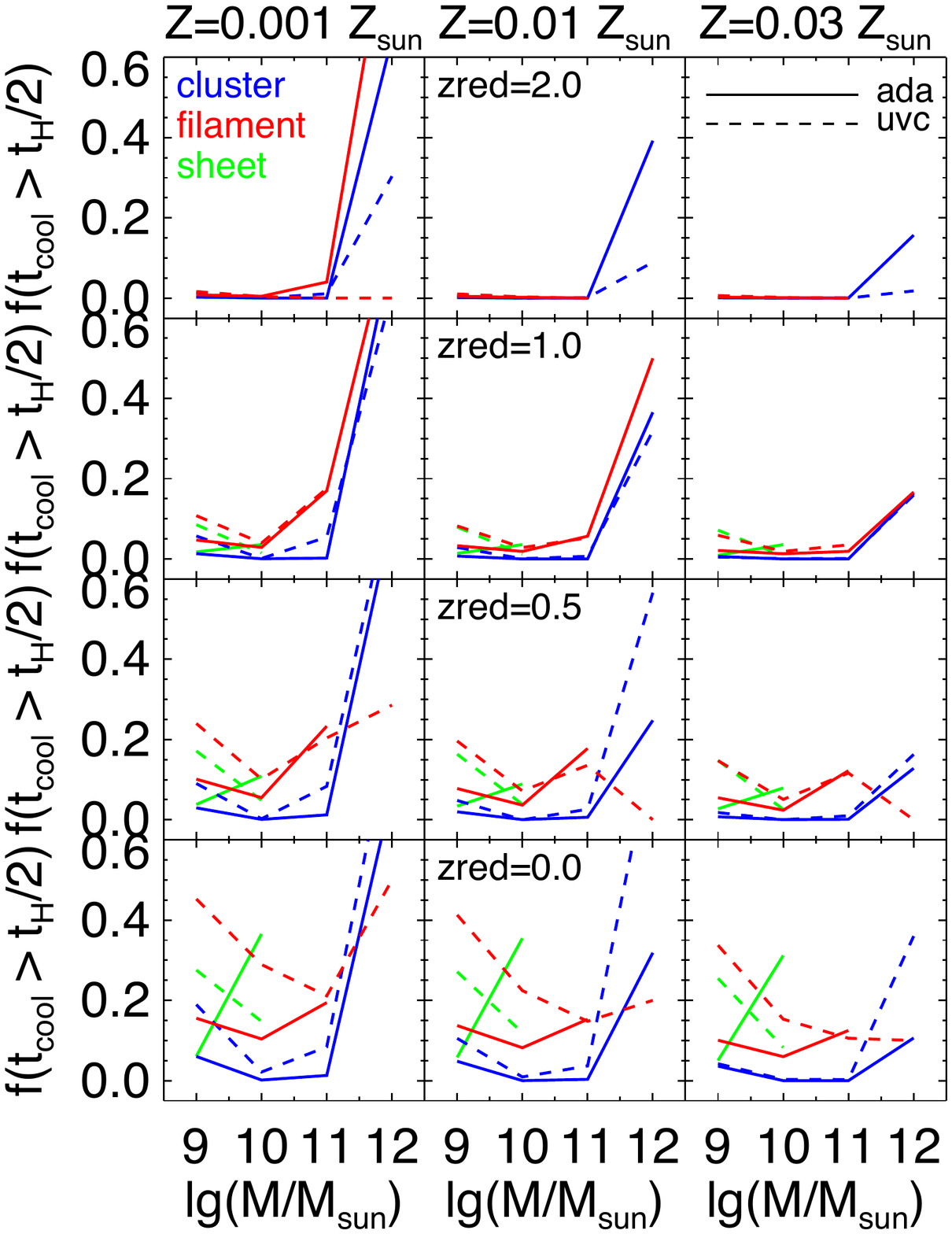}
\end{center}
\vspace{-1.5cm}
\caption{The fractions of halos which have $t_{\rm{cool}}>t_{\rm{H}}$ for gas within the virial radius(Top left), and for gas in $R_{vir}<r<2R_{vir}$(Top right) in clusters(blue), filaments(red) and sheets(green) at redshifts $2.0, 1.0, 0.5, 0.0$; the fractions of halos with $t_{\rm{cool}}>0.5t_{\rm{H}}$ for gas in $R_{vir}<r<2R_{vir}$(Bottom). Gas metallicity $0.001, 0.01, 0.03 Z_{\odot}$ are used respectively to determine $t_{\rm{cool}}$ in the left, middle and right column in each panel. Solid and dashed lines indicate results in simulation L025-ada and L025-uvc respectively. }
\vspace{0.0cm}
\label{figure9}
\end{figure*}

The preheating by gravitational collapse and UV background may delay the cooling and reduce the accretion of baryons in low mass halos. Lu \& Mo(2007) and Lu et al.(2015) investigated the effect of preheated circum-halo gas with a certain entropy level in the semi-analytic preventative models of galaxy formation, and demonstrated that such models can reproduce well a number of scaling relations of galaxies inferred from observations. The gas surrounding low mass halos indeed undergoes evident preheating due to gravitational collapsing and UV background in our simulations. We examine the effect of the preheating in our samples by estimating the cooling time of gas surrounding and within halos, and comparing them to the Hubble time $t_{\rm{H}}$. Following Mo et al.(2005) again, the cooling time reads as

\begin{equation}
\begin{aligned}
t_{\rm{cool}} \sim 6.3 \Lambda_{-23}^{-1}(\frac{\Omega_{B,0}h^2}{0.024})^{-1} (\frac{T}{10^{5.5}\mathbf{\rm{K}}})\\
\times (\frac{1+\delta}{10})^{-1} (\frac{1+z}{3})^{-3} \mathbf{\rm{Gyr}},
\end{aligned}
\end{equation}
where $\Lambda_{-23}$ is the cooling function. The radiative cooling implemented in L025-uvc is calculated with the pristine gas composition during the simulation and following the procedures in Thunes et al.(1998). A more realistic treatment should take the gas metallicity into account. Since the star formation and feedback processes have not been included in our simulation temporarily, we made a post-simulation calculation on the cooling time of circum-halo gas and halo gas, using the cooling functions with metallicities of $0.001, 0.01, 0.03 Z_{\odot}$ given by Sutherland \& Dopita(1993). More specifically, the mean over-density and mass weighted average temperature of gas within $R_{vir}<r<2R_{vir}$, and $r<R_{vir}$ are used to estimate the cooling time of circum-halo gas $t_{cool,cir}$ and halo gas $t_{cool,h}$ respectively. 

The circles representing halos in Figure 6 and Figure 7 are color-coded according to their ratio of the cooling time with metallicity $0.01 Z_{\odot}$ to the Hubble time , i.e., $t_{\rm{cool}}(0.01Z_{\odot})/t_{\rm{H}}$. These two figures suggest that the required average entropy should be around $\sim 2-3$ times of $S_{\rm{pr}}$, i.e., $30-50$\kevcm, in order to reach $t_{\rm{cool}}(0.01Z_{\odot})>t_{\rm{H}}$. In case of gravitational heating being the sole preheating source, only a small fraction of halos in filaments and sheets meets this level of entropy in L025-ada. Adding UV background will raise this fraction. The first and second panel in Figure 9 show the fractions of halos which satisfy $t_{\rm{cool}}>t_{\rm{H}}$ within $r<R_{vir}$, and $R_{vir}<r<2R_{vir}$ in clusters, filaments and sheets since $z=2.0$. We further relax the criterion to $t_{\rm{cool}}>0.5t_{\rm{H}}$, and the fractions are given in the third panel. It is reasonable to expect such a cooling time, $t_{\rm{cool}}>0.5t_{\rm{H}}$, can also delay, at least partially, the gas accretion into halos. 

At $z=2$, a significant delay of cooling occurs only in a few halos residing in clusters with $10^{11.5} M_{\odot}<M_{vir}<10^{12.0} M_{\odot}$. This delay effect for halos in this mass bin persists to low redshifts. The preheating can hardly delay the cooling of circum-halo gas for halos with $10^{10.5} M_{\odot}<M_{vir}<10^{11.5} M_{\odot}$. As for halos with $M_{vir}<10^{10.5} M_{\odot}$, the fraction with $t_{\rm{cool, cir}}>t_{\rm{H}}$ remains only a few percents for all the three metallicity considered at $z \geq1$, and increases to $\sim 5-10 \%$ at $z \lesssim 0.5$ in L025-ada. Generally, the fraction in filaments is the highest, and that in sheets ranks the second for low mass halos, except in $10^{9.5} M_{\odot}<M_{vir}<10^{10.5} M_{\odot}$. However, the number of halos in sheets in this mass bin is much small than that in filaments. The UV heating raises the fraction with $t_{\rm{cool, cir}}>t_{\rm{H}}$ to $\sim 15-30 \%$ at $z \lesssim 0.5$.  A richer metallicity results in a slightly decreased fractions due to more efficient cooling, and vice versa. The fractions of halos meet a relaxed condition, $t_{\rm{cool}}>0.5 t_{\rm{H}}$, at different redshifts are about $\sim 1.3$ times larger. The fractions with $t_{\rm{cool,h}}>t_{\rm{H}}$ are relatively lower than that of $t_{\rm{cool,cir}}>t_{\rm{H}}$ for halos more massive than $10^{10.0} M_{\odot}$. 

\section{Discussion}

\subsection{Comparison with previous studies}

Using the ellipsoidal collapse model, Mo et al. (2005) proposed that the formation of sheets/pancakes can heat the gas surrounding low mass halos to $S \sim 15$\kevcm \, at $z \lesssim 2.0$. They further showed that, with such a level of IGM entropy due to preheating, semi-analytical model of galaxy formation can match the HI mass function and stellar mass function at low mass end simultaneously. In our adiabatic simulation, the gravitational heating due to structures collapse can heat up about $60\%, 45\%$ of the IGM to $S>8, 17$\kevcm \, by $z=0$. However, the fractions drop rapidly toward high redshifts, fell to about $3\%, 1\%$ respectively at $z=2$. A fraction as low as this would limit the power of gravitational preheating to delay the gas cooling at high redshifts. In addition, our analysis suggests that the collapse of filaments, rather than sheets/pancakes, servers as the dominant process of gravitational preheating, contributing more than half of the IGM with $S>8, 17$\kevcm \, since $z=2$. The rapid decrease of gravitation heating toward high redshifts is consistent with the result in Cen(2011)(see their Figure 10 and 11), where the gas entropy were explored in a high resolution simulation run by adaptive grid code ENZO. 

Lu \& Mo (2007) further showed that if the entropy of IGM were preheated up to comparable to or larger than the virail entropy $S_{\rm{vir}}$, the cooling and accretion of gas into dark matter halos can be delayed and reduced. Lu et al.(2015) demonstrated that semi-analytic models with preheated gas entropy $S_{\rm{pr}}$ in the form as eqn(2) can reproduce many observed relations of disk galaxies. In our adiabatic simulation, the entropy of the circum-halo gas of more than $75 \%$ of low mass halos is larger than the virial entropy since $z=2$, due to gravitational preheating. However, the fraction of that larger than $S_{\rm{pr}}$ is merely $\sim 15-20 \%$ for  $M_{\rm{vir}}<10^{10.5} M_{\odot}$ at $z=0$. The fraction of halos whose circum-halo gas having a cooling time longer than $t_{\rm{H}}$ is even less, i.e., grows from a few percents at $z \geq 1$ to $\sim 5-10 \%$ at $z \lesssim 0.5$, by assuming a gas metallicity of $Z \leq 0.03 Z_{\odot}$. Therefore, the gravitational preheating alone is insufficient to sustain the entropy of circum-halo gas urged by the preventative model in Lu et al.(2015).  

The inclusion of UV background as an additional preheating source can almost triple the fractions of halos having $S_{\rm{cir}}>S_{\rm{pr}}$, and of which $t_{\rm{cool, cir}}>t_{\rm{H}}$ for $M_{\rm{vir}}<10^{10.5} M_{\odot}$. However, the combination of gravitational and UV heating is still inadequate to reach the level of gas entropy required in Lu et al.(2015) to reproduce observations. The situation for halos with mass  $10^{10.5}<M_{\rm{vir}}<10^{11.5} M_{\odot}$ would become more serious, as their virial temperature is close to the peak of cooling function at $\sim 10^5\ $K for gas metallicity $Z \leq 0.1 Z_{\odot}$ (Sutherland \& Dopita 1993). Other processes including preheating by supernovae and active galactic nucleus winds (Mo \& Mao 2002) are speculated to offer partly the rest heating energy required by the preventative model in Lu et al.(2015). In Cen (2011) where the star formation and feedback are implemented, the halo gas were found to satisfy $t_{\rm{cool}}>t_{\rm{H}}$ at $z \leq 0.5$ in a considerable fraction of the halos with masses of $10^{10.5}<M_{\rm{vir}}<10^{11.5} M_{\odot}$. However, the preventative feedback alone may be not strong enough to resolve the overcooling in low mass halos, especially for those with masses of $10^{10.5}<M_{\rm{vir}}<10^{11.5} M_{\odot}$. Consequently, the ejective baryonic feedback, which has been vastly adopted and investigated in galaxy formation models and simulations (Naab \& Ostriker 2017 and references therein), will still play an important role. Nevertheless, a precise understanding of the impact of preheating would be helpful in ascertaining the required efficiency and the mechanism of ejective feedback, which vary significantly in various semi-analytical models and simulations at present.

\subsection{Numerical convergence}
As discussed by Mo et al.(2005), the resolutions of shocks, and of low mass halos are crucial to exploring the preheating effect in simulations. Although our simulations are performed on a fixed-grid with the spatial resolution of 24.4 $h^{-1}$ kpc, the advantage of WENO scheme in capturing shocks can resolve well the shocks coming up with the formation of large scale structure in our simulations as demonstrated in Zhu et al. (2013). However, the resolution of dark matter halos in our simulations is largely constrained by the size of grid cell. Underestimated density of gas within and surrounding the halos may amplify the average entropy, i.e., $S_{\rm{cir}}$ and $S_h$, and hence overestimate the preheating effect. This shortcoming is partly remedied by two means in our investigation. First, only halos with more than $250$ dark matter particles are inspected. Second, our investigation put more emphases on the entropy and cooling of gas surrounding the halos, i.e., $R_{vir}<r<2R_{vir}$. 

\begin{figure*}[htbp]
\vspace{-1.0cm}
\hspace{-0.3cm}
\includegraphics[width=0.55\textwidth]{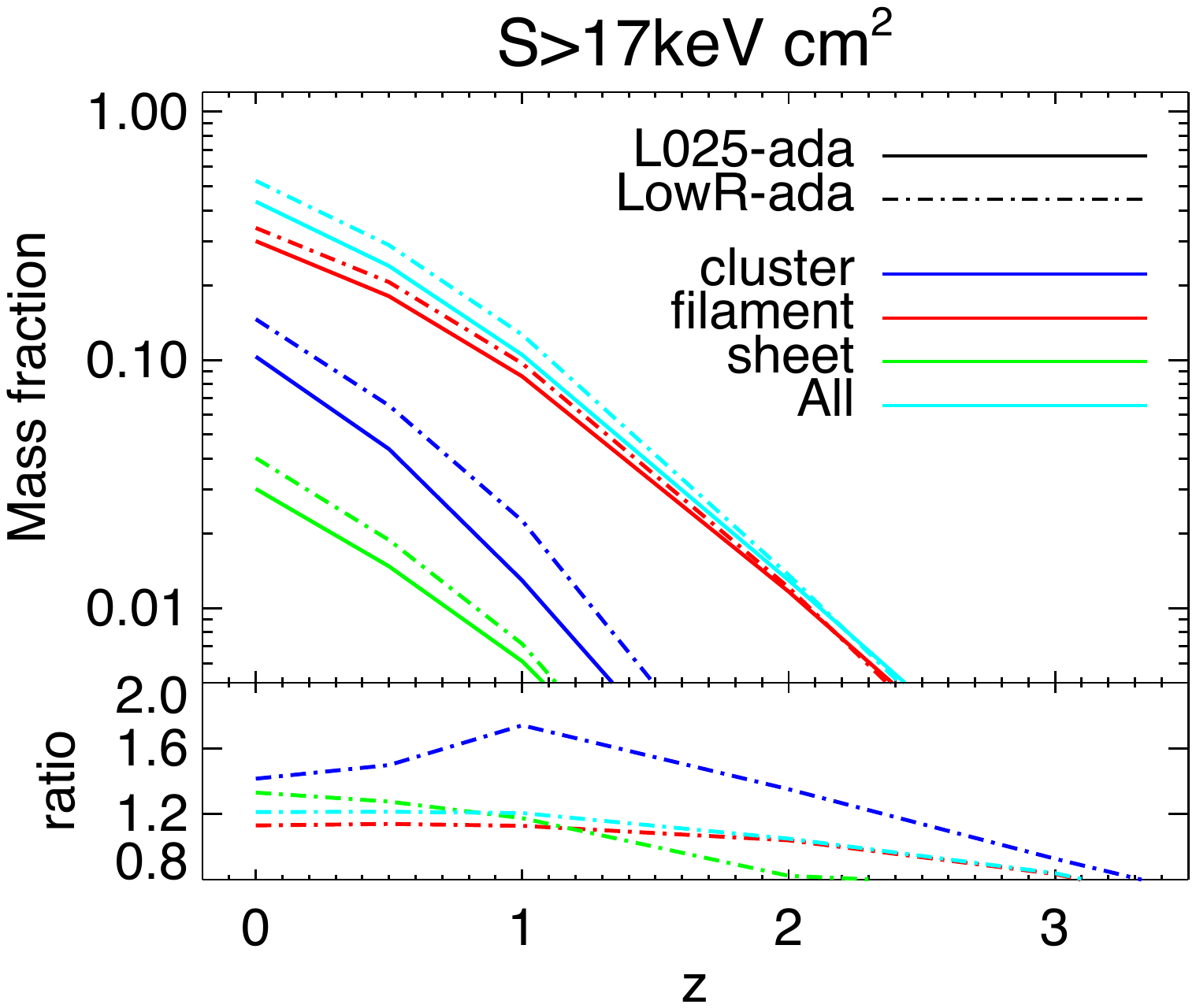}
\hspace{-1.5cm}
\includegraphics[width=0.55\textwidth]{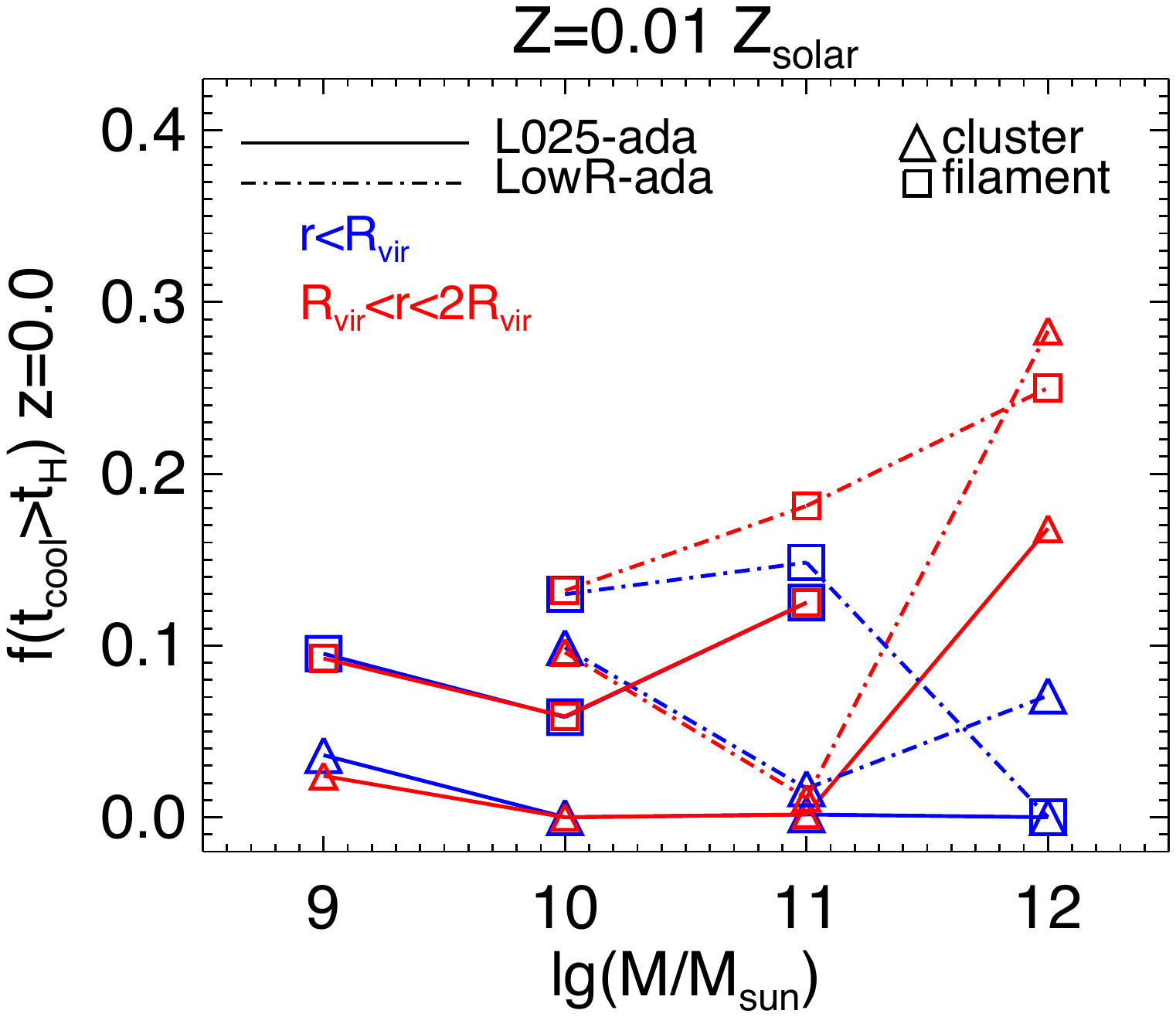}
\vspace{-5.5cm}
\caption{Left: The mass fraction of IGM heated up to $S>17$\kevcm \, in LowR-ada and L025-ada, and the ratios of fractions in LowR-ada to that in L025-ada; Right: The fraction of halos having $t_{\rm{cool}}>t_{\rm{H}}$ with gas metallicity $0.01Z_{\odot}$ in two simulations. The space resolution are 24.4 $h^{-1}$ kpc and 48.8 $h^{-1}$ kpc in L025-ada and LowR-ada respectively. }
\vspace{-0.3cm}
\label{figure10}
\end{figure*}

To test the numerical convergence, we perform an adiabatic simulation with a lower spatial resolution, 48.8 $h^{-1}$kpc, i.e., using a $512^3$ grid and equal number of particles. This simulation is denoted by 'LowR-ada'. Figure 10 shows the comparison between L025-ada and LowR-ada on the fraction of IGM heated up to $S>17 $\kevcm, and the fraction of halos having $t_{\rm{cool}}>t_{\rm{H}}$. Obviously, the contribution to IGM with $S>17 $\kevcm \, from clusters and sheets exhibits a notable discrepancy between two simulations. Nevertheless, minor differences are found regarding the total mass fraction of IGM heated up to $S>17$\kevcm, and those in filaments at $z\leq2.0$. Hence the global effect of gas preheating due to gravitational collapse in two simulations should basically consist with each other. The fractions of halos with  $t_{\rm{cool,cir}}>t_{\rm{H}}$ in L025-ada are smaller than that in LowR-ada by about $5 \%$. Therefore, the insufficiency of gravitational heating to provide the entropy needed by model in Lu et al(2015) might be more serious if the resolution is further increased. 

\subsection{Baryon fraction, environment quenching and large scale conformity}

The delayed cooling by preheating is expected to suppress the accretion rate of gas, and further reduce the star formation and baryon fraction in low mass dark matter halos (Lu \& Mo 2007; Lu et al. 2015). However, the power for estimating the baryon fraction in this work is restrained by the limited resolution of our simulations. In addition, star formation and feedback are also important to the state of baryon in dark matter halos, which are not included yet.  Despite these constraints, the rapid growth of strength of preheating along time at $z<2$ revealed in our simulations is consistent with very recent observation study of the environmental effect on star formation. Based on the Four Star Galaxy Evolution survey, Kawinwanichakij et al. (2017) reported that the environmental quenching was very inefficient at $z>1.5$, but grew rapidly toward low redshifts, and could account for most of the low mass quiescent galaxies($M_{stellar}\sim10^{9-10} M_{\odot}$). The impact of preheating on the baryon fraction of low mass halos will be tackled in the  future by AMR simulations implemented with star formation and feedback.

Moreover, our simulations shows that the filaments is the primary contribution to the preheated IGM with $S> 8$\kevcm. In addition, the fraction of halos with $t_{\rm{cool,cir}}>t_{\rm{H}}$ is relatively higher in filaments than other morphological types of cosmic structures. The rapid growth of this fraction since $z=2$ is actually in accordance to the dramatic rising of filaments during the same epoch (e.g., Zhu \& Feng 2017). Low mass halos in the filaments may witness relatively stronger reduction of gas accretion due to preheating. This trend is supported by some recent observational studies. Using a mass complete sample in the COSMOS field, Darvish et al. (2017) suggested that most satellite galaxies would undergo a rapid quenching as they fall from the field into clusters via filaments. 

On the other hand, the preheating of intergalactic medium could provide a physical explanation for the large scale conformity in gas poor central galaxies that are less massive than $10^{10} M_{\odot}$, although the significancy is still under debate(Kauffman et al. 2013, Kauffman 2015, Berti et al. 2017, Tinker et al. 2017). Our simulations demonstrate that a considerable fraction of the IGM are indeed heated up by the collapsing of large scale structures and the UV ionizing background. So far, these mechanisms are unable to sustain the level of entropy required in semi-analytical models to entirely solve the number density, and scaling relations of low mass galaxies. Whereas, with the circum-halo gas of $\sim 50 \%$ halos going through significant preheating, i.e., $S_{\rm{cir}}>S_{\rm{pr}}$, or $\sim 15 -30 \%$ for $t_{cool,cir}>t_{\rm{H}}$, the preheating sources discussed in this work may be able to cause the reported large scale conformity in the local universe. Further investigation on the impact of preheating on baryon fractions will provide more direct hints. 
 
\section{Summary}
The preheating of IGM, i.e.,heated up to certain entropy levels before collapsing into dark matter halos, have been proposed in preventative models to suppress the formation of low mass galaxies, providing an alternative or a supplement to the ejective models(Mo \& Mao 2002; Mo et al. 2005; Lu \& Mo 2007; Lu, Mo \& Wechsler 2015; Lu et al. 2017). Meanwhile, the preheating of intergalactic medium may offer an physical explanation to the reported large scale galactic conformity(Kauffman et al. 2013, Kauffman 2015). In this paper, we make a study of the preheating of intergalactic medium in three dimensional cosmological hydrodynamical simulations for the first time. We summarise our finding as following: 

(i) Preheating due to pure gravitational collapsing can heat up about $60\%, 45\%$ of the IGM to $S>8, 17$\kevcm \, by $z=0$. However, the fractions drop rapidly as redshift increasing, falling to a few percents at $z=2$, which might be lower than the entropy prescription given by Mo et al.(2005). The inclusion of heating by a uniform UV background (Haardt \& Madau 2012) is unable to increase the fraction of IGM with $S>17$\kevcm. 

(ii) If gravitation collapsing is the sole preheating source, the mass weighted average entropy of the cicum-halo gas $S_{\rm{cir}}$ and halo gas $S_{h}$ in our simulations are higher than the virial entropy for more than $75 \%$ of the halos with $M_{\rm{vir}}<10^{11.5} M_{\odot}$ at $z \leq 2$. However, the fraction of halos whose $S_{\rm{cir}}$ is higher than the entropy level $S_{\rm{pr}}$, required in the preventive model of galaxies formation(Lu et al. 2015), is only $\sim 15-20 \%$ for halos with $M_{\rm{vir}}<10^{10.5} M_{\odot}$ at $z=0$. This fraction drops moderately toward high redshifts. The heating provided by UV background can increase the fraction of $S_{\rm{cir}}>S_{\rm{pr}}$ to nearly $ 70 \%$ for $M_{\rm{vir}}<10^{10.5} M_{\odot}$ at $z=0$, and to above $50 \%$ for $M_{\rm{vir}}<10^{9.5} M_{\odot}$ at $z>0$. 

(iii) Assuming a metallicity $Z \leq 0.03 Z_{\odot}$, we measure the fraction of halos whose circum-halo gas would experience significant delay of cooling, i.e. having a cooling time longer than the Hubble time $t_{\rm{cool,cir}}>t_{\rm{H}}$.  As for halos less massive than $10^{10.5} M_{\odot}$, the fraction of halos that have $t_{\rm{cool,cir}}>t_{\rm{H}}$ remains only a few percents at $z \geq 1$, and increases to $5-10 \%$ at $z \lesssim 0.5$ due to gravitational heating. The fractions of halos that meet a relaxed criterion, $t_{\rm{cool,cir}}>0.5 t_{\rm{H}}$ are about $\sim 1.3$ times larger.  A richer metallicity results in a slightly decreased fractions and vice versa. The UV heating can raise the fraction of halos with $t_{\rm{cool,cir}}> t_{\rm{H}}$ to $\sim 15-30 \%$ at $z \lesssim 0.5$. However, hardly any halos with masses $10^{10.5}-10^{11.5} M_{\odot}$ will be affected, as their virial temperature is close to the peak of cooling function at $\sim 10^5$K and UV heating can hardly help. The cooling of halo gas is generally more efficient than the circum-halo gas for halos more massive than $10^{10.0} M_{\odot}$. 

(iv) The formation of filaments, instead of sheets/pancakes, servers as the primary gravitational preheating process, contributing more than half of the IGM with $S>8, 17\ $\kevcm. In addition, the fraction of halos with $t_{\rm{cool,cir}}>t_{\rm{H}}$ in filaments is generally the highest, and that in sheets ranks the second in the same halo mass bin. This trend is demonstrated both in simulations with and without including the UV background. 

The results presented in this paper indicate that preheating due to the collapse of cosmic structures and UV background can provide part of the IGM entropy required by semi-analytical preventative model in order to match the observed properties of low mass galaxies, such as the number density, and scaling relation and so on. To fulfil the request by preventative models, more sources are urged to heat the gas surrounding low mass dark halos, especially those with mass $10^{10.5}-10^{11.5} M_{\odot}$. Nevertheless, these two preheating mechanisms explored in this work might be able to give rise to the observed large scale conformity at local universe. The validity will be testified by further investigation regarding the impact of preheating on the baryon fraction and star formation in low mass halos, combining with more observations in the future. 

\begin{acknowledgements}
Acknowledgements: 
\end{acknowledgements}
The authors thanks the anonymous referee for his helpful comments to improve the manuscript. This work was partly supported by the Key Project of the National Natural Science Foundation of China under(NSFC) 11733010, and National Key Research and Development Plan of China(No. 2017YFB0203302). W.S.Z. is supported by the NSFC grants 11673077 and the Fundamental Research Funds for the Central Universities. F.L.L. is supported under the NSFC grants 11333008, and State Key Development Program for Basic Research of China (Nos. 2013CB834900 and 2015CB855000)

\appendix
\section{mass fractions of IGM in various phases}

Figure A1. gives the mass fractions of IGM in four types of density-temperature phases in our simulations. The definition of IGM phases varies in the literature. Throughout this paper, the definition in Dave et al(2010) with the thresholds of $T_{th}=10^5$ K and $\delta_{th}=100$ are adopted, i.e., diffuse($T<T_{th}, \delta_b<\delta_{th}$), WHIM ($T>T_{th}, \delta_b<\delta_{th}$), hot halo ($T>T_{th}, \delta_b>\delta_{th}$), and condensed ($T<T_{th}, \delta_b>\delta_{th}$). Due to the lack of star formation process, the mass fraction of stellar component is not available in our simulations. The global trend and values of the fractions in L025-uvc are generally in agreement with Dave et al.(2010). Along with the growth of cosmic structures, gas in the diffuse phase  keep converting into WHIM, hot halo and condensed phases. The mass fraction of diffuse gas drops from $\sim 90\%$ at $z=3$ to $\sim 45\%$ at $z=0$. Meanwhile, the fractions of gas in the WHIM and hot halo grows from a few percents to $ \sim 20\%$. The condensed baryons contributes about $10 \%$ since $z=3$. We note that the mass fractions in the phases of diffuse, WHIM and hot halo gas extracted from L025-uvc show differences of $\sim 5\%$ in absolute terms comparing with Dave et al.(2010) at $z \leq 0.5$. While similar differences can be found between various works (e.g., Cen et al. 2006; Dave et al. 2010; Snedden et al. 2016), this discrepancy can be attributed to different modeling of physical processes including the star formation, supernova and stellar feedback , as well as some numerical factors including the hydrodynamic solver and resolution.

\begin{figure}[htbp]
\vspace{-1.5cm}
\begin{center}
\includegraphics[width=0.55\textwidth]{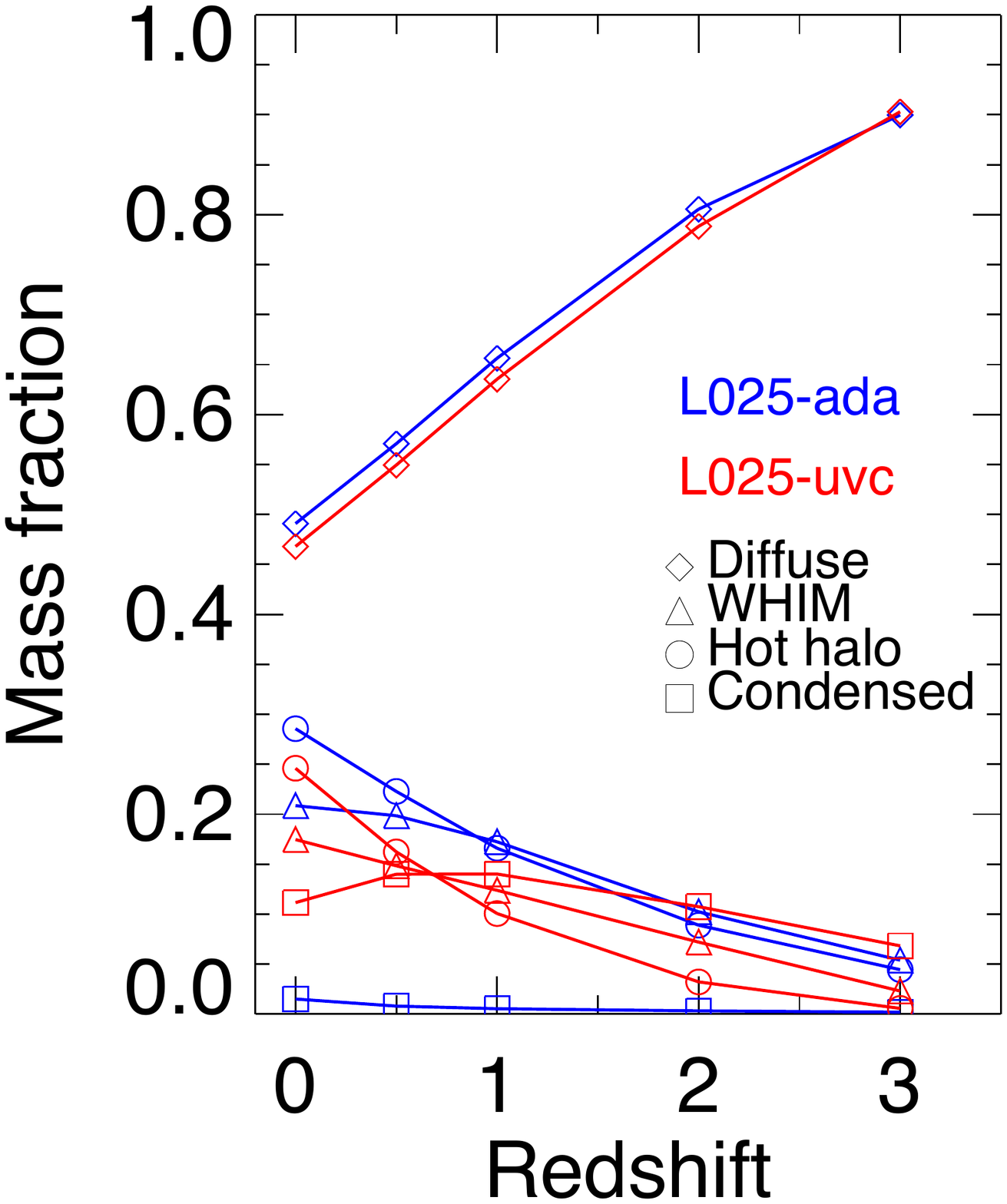}
\end{center}
\vspace{-1.5cm}
\caption{The mass fractions of IGM in various density-temperature phases.}
\vspace{-0.3cm}
\label{figure-A1}
\end{figure}

\end{document}